\documentclass[12pt, final]{ucdavisthesis}

\usepackage[us,nodayofweek,12hr]{datetime}
\usepackage{graphicx}
\usepackage{hyperref}
\usepackage[affil-it]{authblk}
\usepackage{amsmath, amsfonts, amsthm, amssymb}
\usepackage[usenames,dvipsnames]{color}
\usepackage{comment}
\usepackage{ifpdf}
\usepackage{bbold}
\usepackage[utf8]{inputenc}
\usepackage[english]{babel}
\usepackage{graphicx}  
\usepackage{breqn}
\usepackage{mathrsfs}
\usepackage{amsmath}
\usepackage[affil-it]{authblk}
\usepackage{blindtext}
\usepackage{amsmath, amsfonts, amsthm, amssymb}
\usepackage[usenames,dvipsnames]{color}
\usepackage{subcaption}
\usepackage{url}
\usepackage{comment}
\usepackage{ifpdf}
\usepackage{bbold}
\usepackage[utf8]{inputenc}
\usepackage{graphicx}  
\usepackage{breqn}
\usepackage{mathrsfs}
\usepackage{cite}
\usepackage[toc,page]{appendix}
\newcommand*\colvec[3][]{
    \begin{pmatrix}\ifx\relax#1\relax\else#1\\\fi#2\\#3\end{pmatrix}
}

\title          {A Tour Through Shape Dynamic Black Holes}

\author         {Gabriel Herczeg}

\authordegrees  {B.S. Physics (Brooklyn College) 2010 \\
		       B.S. Math (Brooklyn College) 2010 \\
                 M.S. Physics (University of California, Davis) 2012}

\officialmajor  {Physics}

\graduateprogram{Physics}

\degreeyear     {2017}

\degreemonth    {June}

\committee{Professor Steven Carlip}{Professor Markus Luty}{Professor Andrew Waldron}{}{}

\dedication{\textsl{To Bunny \ldots\\ \vspace{4mm}
For never letting me lose sight of what really matters: \\ Making good memories with family and good friends. \\
\vspace{4mm} And to my family and good friends \ldots  you know who you are.
}}

\abstract{\noindent Shape dynamics is a classical theory of gravity which agrees with general relativity in many important cases, but possesses different gauge symmetries and constraints. Rather than spacetime diffeomorphism invariance, shape dynamics takes spatial diffeomorphism invariance and spatial Weyl invariance as the fundamental gauge symmetries associated with the gravitational field. Despite these differences, shape dynamics and general relativity generically predict the same dynamics---there exist gauge-fixings of each theory that ensure agreement with the other. However, these gauge-fixing conditions are not necessarily globally well-defined and it is therefore possible to find solutions of the shape dynamics equations of motion that agree with general relativity on some open neighborhoods, but which have different global structures. In particular, the black hole solutions of the two theories disagree globally. Understanding these novel ``shape dynamic black holes" is the primary goal of this thesis.}

\acknowledgments{ \begin{itemize}
\item
General Acknowledgments: \\
I am extremely grateful to Steve Carlip for encouraging this research, and for providing outstanding guidance and mentorship throughout my time in Davis. I am also very grateful to Henrique Gomes for welcoming me into the very impressive group of scientists that comprise the shape dynamics community. Many thanks to Markus Luty and Andrew Waldron whose helpful suggestions and thoughtful, constructive criticism have improved this manuscript.  
\item
Chapter 2: \\
This chapter was based on a collaboration with Henrique Gomes. I would like to thank Steve Carlip for his support and for his many helpful questions, comments and constructive criticisms. I would also like to thank Vasudev Shyam and Joshua Cooperman for their valuable input and insightful questions about this and related research. H.G. was supported in part by the U.S.Department of Energy under grant DE-FG02-91ER40674.
\item
Chapter 3: \\
This chapter was based on a collaboration with Vasudev Shyam. I would like to thank Henrique Gomes for helpful conversations. I am also grateful to Sean Gryb for lively discussions and for his generous hospitality during the final stages of this work, and to Flavio Mercati for helpful suggestions. Many thanks to Steve Carlip for his valuable insights and suggestions. Research for this chapter was supported in part by FQXi minigrant MGA-1404 courtesy of the Foundational Questions Institute. V.S. was supported in part by Perimeter Institute for Theoretical Physics. Research at Perimeter Institute is supported by the Government of Canada through Industry Canada and by the Province of Ontario through the Ministry of Research and Innovation.
\item
Chapters 4-6: \\
I am grateful to Steve Carlip for carefully reading drafts and making many helpful suggestions, and to Joseph Mitchell for many lively conversations about this work as it progressed. I would also like to thank Jack Gegenberg for helpful discussions in the early stages of this work, and Henrique Gomes, Sean Gryb, Flavio Mercati and Vasudev Shyam for their thoughtful questions and remarks.
\end{itemize}
}

\begin{document}

\newcommand{\overbar}[1]{\mkern 1.5mu\overline{\mkern-1.5mu#1\mkern-1.5mu}\mkern 1.5mu}
\newcommand{\pder}[2]{\frac{\partial#1}{\partial#2}}
\newcommand{\bibfont}{\singlespacing}

\bibliographystyle{ieeetr}
\makeintropages 

\chapter{Canonical General Relativity}
\pagenumbering{arabic}

One of the most celebrated acheivements of twentieth century physics was the description of gravity in terms of the curvature of spacetime. The general theory of relativity was completed in the year 1915 and in the very same year David Hilbert proposed an action integral reproducing the Einstein field equations:

\begin{equation}\label{Einstein-Hilbert}
S_{\mbox{\tiny EH}} = \frac{1}{16\pi G_N}\int_{M} d^4x \sqrt{-g}R[g]
\end{equation}

\noindent where $G_N$ is Newton's constant, $g$ is the spacetime metric, $R[g]$ is the scalar curvature of $g$, and units have been chosen such that the speed of light $c=1$. The integral is carried out over the entire spacetime manifold $M$.

Despite the fact that a variational principle was identified very early in the development of general relativity, it would not be until the year 1959 that Richard Arnowitt, Stanley Deser and Charles Misner (ADM) were first able to reformulate general relativity as a constrained Hamiltonian system \cite{ADM-first}. Since then, their original paper has received thousands of citations and has paved the way for many important advances in classical and quantum gravity. Some important examples include numerical relativity; the Wheeler-Dewitt equation \cite{Wheeler}; Ashtekar's  formulation of general relativity in terms of ``new variables" \cite{AshtekarNew} and loop quantum gravity \cite{JacobsonSmolin, RovelliSmolin, PullinLewandowski, RovelliSpin}; Euclidean quantum gravity \cite{EuclideanQuantumGravity}; causal dynamical triangulations \cite{CDT}; Ho\v{r}ava-Lifshitz gravity \cite{Horava} and recently, shape dynamics \cite{GGK} to name just a few. 

The key insight provided by ADM was that any arbitrarily chosen local time coordinate $t$ defines a local foliation of spacetime by space-like hypersurfaces $\Sigma_t$ of constant $t$, often called ``time slices." The spacetime metric $g_{\mu\nu}$ can then be decomposed in terms of the spatial metric $q_{ij}$ induced on the time slices, a three-vector $\xi^i$ on the spatial slices called the shift vector, and a scalar function $N$ called the lapse function. 

\begin{figure}[h]
\begin{center}
\includegraphics[width = 11.5cm, height = 8.5cm]{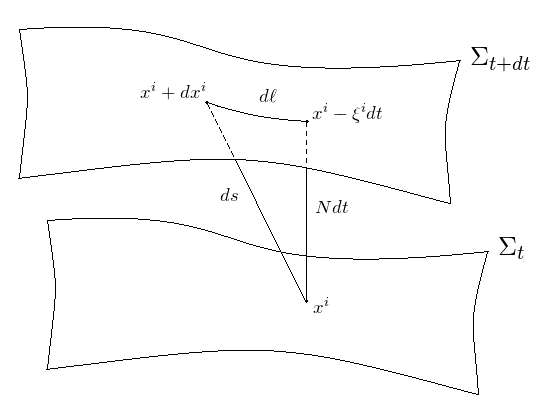} \\
\end{center}
\caption{Two time slices separated by an infinitesimal change $dt$ in the time coordinate $t$. The lapse $N$ measures the infinitesimal change along the direction normal to the hypersurfaces, while the shift $\xi^i$ measures the change in the spatial coordinates $x^i$ produced by moving in the normal direction.}
\end{figure}

In terms of this decomposition, one can think of the spacetime line element $ds$ as the hypotenuse of an infinitesimal ``Lorentzian right triangle" with one leg $Ndt$ orthogonal to the time slices and one leg $d\ell$ (whose length is measured relative to the induced metric $q_{ij}$) lying tangent to the later time slice $\Sigma_{t+dt}$ This leads to:
\begin{equation}\label{ADM}
ds^2 = -(Ndt)^2 + q_{ij}\left(dx^i+\xi^idt\right)\left(dx^j+\xi^jdt\right).
\end{equation}

Comparing \eqref{ADM} to the standard exression for the spacetime line element $ds^2 = g_{\mu\nu}dx^{\mu}dx^{\mu}$, one can read off the components of the spacetime metric in terms of the ADM variables.
\begin{subequations}
\label{ADMall}
\begin{align}
g_{00} &= -N^2 + \xi^i\xi^j q_{ij} \label{ADM1} \\
g_{0i} &= g_{i0} = q_{ij}\xi^j \label{ADM2}\\
g_{ij} &= q_{ij}. \label{ADM3}
\end{align}
\end{subequations}

\noindent One can also display \eqref{ADMall} as a matrix equation
\begin{equation}
\label{gmatrix}
g =
\left(\begin{array}{ccc}
-N^2 + \xi^m \xi_m & \xi_i \\
 \xi_j & q_{ij}\\
\end{array} \right)
\end{equation}

\noindent which has the inverse
\begin{equation}
\label{g_inv}
g^{-1} =
\frac{1}{N^2} \left(\begin{array}{ccc}
-1 & \xi^k \\
\xi^i & N^2q^{ik}- \xi^i\xi^k\\
\end{array} \right)
\end{equation}

\noindent where the spatial metric was used to lower indices of the spatial shift vector as $\xi_i = q_{ij}\xi^j$.

Ordinarily, one thinks of the Einstein-Hilbert action \eqref{Einstein-Hilbert} as being a functional of the spacetime metric $g_{\mu\nu}$, but making use of \eqref{ADMall} it can just as easily be thought of as a functional of $N$, $\xi^i$ and $q_{ij}$. In terms of the ADM decomposition, the action \eqref{Einstein-Hilbert} becomes
\begin{equation}
\label{ADM Action 1}
S_{\mbox{\tiny EH}} = \frac{1}{16\pi G_N}\int dt \int_{\Sigma_t} d^3x \sqrt{q}N\left(R[q] + K_{ij}K^{ij} - K^2 \right)
\end{equation}

\noindent where $R[q]$ is the scalar curvature of $q_{ij}$, $K_{ij}$ is the extrinsic curvature of $\Sigma_t$ embedded in $M$, defined by:
\begin{subequations}
\label{extrinsic}
\begin{align}
K_{ij} &:= \mathcal{L}_{n}g_{ij} \label{exDef} \\
&\,\,= \frac{1}{2N}\left(\dot{q}_{ij} - \nabla_i\xi_j - \nabla_j\xi_i\right) \label{exEquals} 
\end{align}
\end{subequations}

\noindent and $K=q_{ij}K^{ij}$ is the trace of the extrinsic curvature. In \eqref{extrinsic} $\mathcal{L}_X$ is the Lie derivative along the vector field $X$, $n^{\mu} = Ng^{t\mu}$ is the unit normal vector to the time slice $\Sigma_t$, and $\nabla_i$ is the Levi-Civita connection associated with $q_{ij}$. 

One can now compute the momenta conjugate to the ADM variables:
\begin{eqnarray}
\pi^{ij} &=& \frac{\delta S}{\delta \dot{q}_{ij}} = \sqrt{q}\left(K^{ij}-q^{ij}K\right) \\
\pi^0 &=& \frac{\delta S}{\delta \dot{N}} = 0 \\
\pi^i &=& \frac{\delta S}{\delta \dot{\xi}_i} = 0
\end{eqnarray}

The momenta $\pi^0$ and $\pi^i$ conjugate to the Lagrange multipliers $N$ and $\xi_i$ vanish and are therefore primary constraints. The equation defining the momentum $\pi^{ij}$ conjugate to $q_{ij}$ is equivalent to
\begin{equation}
K^{ij} = \frac{1}{\sqrt{q}}\left(\pi^{ij}-q^{ij}\pi\right)
\end{equation}

\noindent where $\pi = \pi^{ij}q_{ij}$. The action can now be rewritten \eqref{Einstein-Hilbert} in terms of the constraints and canonical variables:
\begin{eqnarray}
S_{\mbox{\tiny EH}} &=& \frac{1}{16\pi G_N}\int dt\int_{\Sigma}d^3 x \left(\pi^{ij}\dot{q}_{ij} - N\mathcal{S} - \xi^i\mathcal{H}_i\right) \\
\mathcal{S} &=& \frac{1}{\sqrt{q}}\left(\pi_{ij}\pi^{ij} - \pi^2\right) - \sqrt{q}R[q] \\
\mathcal{H}_i &=& -2\nabla_j\pi^j_{\,\,i}.
\end{eqnarray}

Having performed the Legendre transformation, the canonical Poisson bracket relations can now be read from from the symplectic potential term $\dot{q}_{ij}\pi^{ij}$:
\begin{eqnarray}
\label{PB}
\{q_{ij}(x),\pi^{kl}(y)\} &=& 16\pi G_N \delta^k_{(i}\delta^l_{j)}\delta^3(x-y) \\
\{q_{ij}(x),q_{kl}(y)\} &=& 0 \\
\{\pi^{ij}(x),\pi^{kl}(y)\} &=& 0 
\end{eqnarray}

The remainder of the Legendre transformed Lagrangian $N\mathcal{S}+\xi^i\mathcal{H}_i$ is identified as the gravitational Hamiltonian, which is pure constraint. From the canonical commutation relations \eqref{PB}, one can compute the brackets of the constraints with each other:
\begin{subequations}
\label{C-algebra}
\begin{eqnarray}
\{\mathcal{S}(N_1),\mathcal{S}(N_2)\} &=& \mathcal{H}_i(q^{ij}\left(N_1\partial_iN_2-N_2\partial_iN_1)\right) \label{bad1} \\
\{\mathcal{S}(N),\mathcal{H}_i(\xi^i)\} &=&  \mathcal{S}(\xi^i\partial_iN)\\
\{\mathcal{H}_i(\xi_1^i),\mathcal{H}_j(\xi_2^j)\} &=& \mathcal{H}_k([\xi_1,\xi_2]^k).
\end{eqnarray}
\end{subequations}

\noindent where I have adopted units such that $16\pi G_N = 1$. It is evident from \eqref{C-algebra} that the constraints form a closed algebra, and are therefore said to be ``first-class." However, it is also clear that the right hand side of \eqref{bad1} depends on the inverse spatial metric $q^{ij}$---thus, the algebra of constraints is said to have ``structure functions" rather than structure constants, since the algebra of constraints varies from point to point in phase space \cite{Wald-Lee}. The constraint algebra is therefore not an ordinary Lie algebra---a fact which has been a stumbling block toward understanding canonical general relativity in standard gauge-theoretic terms. It will be shown in the following chapter that shape dynamics avoids this complication.
\chapter{Shape Dynamics}
\section{What is Shape Dynamics?}

Shape dynamics is a classical, Hamiltonian theory of gravity in which
solutions are described by the time evolution of spatial
(three-dimensional Riemannian) conformal geometries
\cite{GGK} rather than spacetime geometries as in general relativity. Remarkably, while the two theories have different physical degrees of freedom and different gauge redundancies, the dynamics of the two theories agree in a broad range of circumstances.

While there are certainly some great advantages to the general relativistic description of gravity over shape dynamics, most notably that general relativity has a local Hamiltonian and a clear description of test particle motion in the limit of no back-reaction relying solely on the geometry of spacetime, shape dynamics with its peculiar non-local Hamiltonian has some advantages as well. The pure-constraint structure of ADM Hamiltonian has the well-known problem that it entangles dynamics with gauge transformations for the simple reason that time-evolution can be achieved by performing a transformation of the time variable. On the other hand, shape dynamics has a vastly restricted class of allowable time variables, leading to a clear separation between gauge transformations and physical time-evolution. Additionally, shape dynamics has the benefit that the constraint algebra does not vary from point to point of phase space---it is a bona-fide Lie algebra. Both of these features have raised hopes that the shape dynamics description of gravity might be more amenable to canonical quantization than the general relativistic description, although it must be admitted that these hopes are so far largely unfulfilled. 

The canonical variables of shape dynamics are given by a  Riemannian
metric $q_{ij}$ and its conjugate momentum $\pi^{ij}$. It is
instructive to compare shape dynamics with the canonical formulation
of general relativity due to Arnowitt, Deser and Misner (ADM) \cite{ADM} summarized in the preceding chapter. In the ADM formalism, the gauge symmetries are spatial diffeomorphism and (on--shell) refoliation invariance. Refoliation invariance allows one to transform from a solution on one family of spacelike hypersurfaces to a physically equivalent solution on another. Shape dynamics does not possess refoliation invariance; the solutions are described by the evolution of a conformal class of three-dimensional Riemannian geometries with respect to a time \emph{parameter} that is fixed (up to monotonic reparemetrizations $t \to f(t)$). 

Rather than refoliation invariance, the equations of motion of shape dynamics are invariant under local scale transformations of the spatial metric. The result is a theory which possesses different gauge symmetries than general relativity, but which is nevertheless generically \emph{locally} equivalent to it. Local equivalence holds in the sense that around a generic point in a solution to Einstein's equations there is a local patch which can  be directly mapped onto a shape dynamics solution, and vice-versa. The mapping is achieved by simultaneously applying (partial) gauge fixings on each theory \cite{Linking}.  

Local scale transformations (here called spatial Weyl transformations)
are generated by the trace of the momentum $\pi = \pi^{ij}
q_{ij}$. The statement that spatial Weyl transformations are a gauge
symmetry of shape dynamics\footnote{The full set of Weyl
  transformations are easier to implement in case one has a
  non-closed spatial manifold. In case the manifold is closed, shape
  dynamics is constructed with the group of Weyl transformations that
  preserve the total volume of space.} is translated in the Dirac
constraint formalism  into $\pi = 0$ being a first class constraint on
the phase space of shape dynamics. It is only for solutions of ADM for
which one can find a global foliation in maximal slicing--  i.e. that
for which $\pi=0$ on each hypersurface--  that ADM can be dual to shape dynamics. For certain important cases, such as the ones investigated in this thesis, this is not the case. 

{For this reason the notion of a line element is carefully distinguished from that of a spacetime metric, the former being viewed as an ansatz to construct the latter which is successful only when the line element is non-degenate. This is because shape dynamics does not possess spacetime diffeomorphism invariance, as I have stressed above. A reconstructed solution always exists and gives rise to a line element in the maximal foliation, but in many circumstances the line element will not form a non-degenerate, or a non-singular, spacetime metric.} 


The most straightforward way to construct shape dynamics is to make use of a tool called the ``linking theory" that serves as a theoretical bridge between the ADM formulation of general relativity and shape dynamics. One starts with the ADM phase space, and then extends this phase space by a new pair of conjugate variables $(\phi,\pi_{\phi})$. In more concrete terms, one begins with the ADM action, and perform the canonical embedding into the phase space of the linking theory defined by
\begin{align}
q_{ij} \to t_{\phi}(q_{ij}) := e^{4\phi}q_{ij} \label{embed1} \\
\pi^{ij} \to t_{\phi}(\pi^{ij}) := e^{-4\phi}\pi^{ij} \label{embed2}
\end{align}

\noindent where $q_{ij}$ is the spatial metric induced on a constant-$t$ hypersurface, and $\pi^{ij}$ is its conjugate momentum. This procedure yields the action of the linking theory, which now possesses not only the transformed ADM constraints which generate on-shell refoliations and spatial diffeomorphisms, but also a new ``conformal constraint" which generates spatial Weyl transformations. In order to obtain shape dynamics, one then imposes the gauge fixing condition $\pi_{\phi} = 0$, which reproduces the original phase space $(q_{ij},\pi^{ij})$. The resulting theory depends on the field $\phi$, but $\phi$ is no longer a dynamical variable. One can show that as a result of performing this phase space reduction, the scalar constraint $\mathcal{S}(x)$ is no longer first class with respect to the diffeomorphism constraint $H_a(x)$ and the conformal constraint $\pi(x)$ which are the only remaining first class constraints in shape dynamics. The result is a theory which, loosely speaking, is defined on a fixed maximal foliation, where the conformal constraint $\pi(x) = 0$ can be identified with the maximal slicing condition commonly used for finding valid initial data in the ADM formulation of general relativity. What distinguishes shape dynamics from ADM general relativity in maximal slicing is that the role of the gauge fixing conditions and constraints have been reversed---$\pi(x)$ is now viewed as a first class constraint generating spatial Weyl transformations rather than as a gauge-fixing condition on the ADM phase space. For more details on the construction and historical development of shape dynamics see \cite{GGK, poincare, Linking, tutorial, superspace}. It is worth noting that the form of the constraints and equations of motion for shape dynamics differ somewhat depending on the spatial topology. Throughout this paper asymptotically flat boundary conditions are assumed on a spatially non-compact topology as in \cite{poincare}. For details on the spatially compact case see e.g. \cite{GGK}. Readers seeking additional background on shape dynamics should refer to \cite{GGK, H-thesis, tutorial, FAQ, York, gravDOF, TimMatter}

\section{Construction of Shape Dynamics from the Linking Theory}
Before continuing to the main results concerning shape dynamic black holes, it is worthwhile to review the construction of shape dynamics for asymptotically flat spatial manifolds presented in \cite{H-thesis} and elaborated in \cite{poincare}. One begins with the standard canonical formulation of general relativity. The canonical form of the Einstein-Hilbert action is given by
\begin{eqnarray}\label{ADM action}
\mathcal{I}_{\mbox{\tiny ADM}} &=& \frac{1}{16\pi}\int dt \int_{\Sigma} d^3x\sqrt{q} \left(\dot{q}_{ij}\pi^{ij} -
  N\mathcal{S}(x) - \xi^i\mathcal{H}_i(x)\right) \nonumber \\  &-&
\frac{1}{8\pi}\int dt \int_{\partial\Sigma} d^2x\sqrt{\sigma}(NK+r^iN_{,i}-r_i\xi^j\pi^i_j)
\end{eqnarray}

\noindent where $\pi^{ij} = \frac{\partial\mathcal{L}}{\partial \dot{q}_{ij}}$ is the momentum  canonically conjugate to $q_{ij}$, $\sigma_{ab}$ is the metric induced on $\partial\Sigma$ by $q_{ij}$, $r^i$ is the outward pointing normal vector of  $\partial\Sigma$, $K$ is the trace of the extrinsic curvature of $\partial\Sigma$ embedded in $\Sigma$, and $N$ and $\xi^i$ are the lapse function and shift vector. The quantities $\mathcal{S}(x)$ and $\mathcal{H}_a(x)$ are the scalar constraint (or ``Hamiltonian constraint") and the diffeomorphism constraint (or ``momentum constraint"), defined by \eqref{C-algebra}.

The scalar constraint $\mathcal{S}(x)$ generates refoliations of spacetime provided the equations of motion are satisfied, and the diffeomorphism constraint $\mathcal{H}_i(x)$ generates foliation-preserving spatial diffeomorphisms.

The next step in the construction of shape dynamics is to extend the phase space of canonical general relativity by a scalar field $\phi$ and its canonically conjugate momentum $\pi_{\phi}$, where $e^{4\phi}$ plays the role of a ``conformal factor" satisfying the fall-off condition $e^{4\phi} \sim 1 + \mathcal{O}(r^{-1})$. Together with the new first class constraint $\pi_{\phi} \approx 0$, the addition of these new canonical variables trivially embeds canonical general relativity into the extended phase space $(q_{ij}, \pi^{ij},\phi, \pi_{\phi})$. 

To obtain a non-trivial embedding, one then performs the canonical transformation 
\begin{eqnarray}
T_{\phi}q_{ij}(x) &:=& e^{4\phi(x)}q_{ij}(x) \nonumber \\
T_{\phi}\pi^{ij}(x) &:=& e^{-4\phi(x)}\pi^{ij}(x) \nonumber \\
T_{\phi}\phi(x) &:=& \phi(x) \nonumber \\
T_{\phi}\pi_{\phi}(x) &:=& \pi_{\phi}(x)-4\pi(x) \nonumber
\end{eqnarray}

\noindent The Hamiltonian for the extended phase space can be written in terms of the canonically transformed first class constraints:
\begin{equation}\label{H-linking}
\mathcal{H}_{\mbox{\tiny link}} = \int_{\Sigma}d^3x\left[N(x)T_{\phi}\mathcal{S}(x) + \xi^i(x)T_{\phi}\mathcal{H}_i(x) + \rho(x)T_{\phi}\pi_{\phi}(x)\right] + T_{\phi}B(N,\xi)
\end{equation}

\noindent where $N(x)$ is the lapse function, $\xi^a(x)$ is the shift vector, $\rho(x)$ is a Lagrange multiplier for the Weyl constraint $T_{\phi}\pi_{\phi}$, $B(N,\xi)$ is the Gibbons-Hawking-York boundary term (the surface integral in \eqref{ADM action}), and $T_{\phi}B(N,\xi)$ plays an important role in defining the globally conserved charges in shape dynamics \cite{poincare}. The system described by this Hamiltonian is known as the \emph{linking theory}, as it provides a link between canonical general relativity and shape dynamics \cite{Linking}. In order to obtain shape dynamics from the linking theory, one imposes the gauge fixing condition $\pi_{\phi}(x) \equiv 0$. The only constraint whose Poisson bracket with this condition is weakly non-vanishing is $T_{\phi}\mathcal{S}(N)$:
\begin{equation}\label{weak bracket}
\{T_{\phi}\mathcal{S}(N),\pi_{\phi}(x)\} = 4T_{\phi}\{\mathcal{S}(N),\pi(x)\},
\end{equation}

\noindent where $\mathcal{S}(N) = \int_{\Sigma}d^3xN(x)S(x)$ is the scalar constraint smeared with the lapse function $N(x)$. Equation \eqref{weak bracket} implies the so-called lapse-fixing equation\footnote{The quantity $G_{ijkl}$ appearing in the lapse fixing equation is the DeWitt supermetric defined by: \\ $G_{ijkl}:= \frac{1}{2}(q_{ik}q_{jl} + q_{il}q_{jk})-q_{ij}q_{kl}.$ \\}:
\begin{equation}\label{LFE1}
T_{\phi}\{\mathcal{S}(N), \pi(x)\} \approx
e^{-4\phi}(\nabla^2 N + 2q^{ij}\phi_{,i}N_{,j}) -
e^{-12\phi}NG_{ijkl}\frac{\pi^{ij}\pi^{kl}}{|q|} \approx 0,
\end{equation}

\noindent where $R$ is the scalar curvature of $q_{ij}$, and $\approx$ denotes ``weak equality"---i.e. equality up to an additive term that vanishes when the constraints are satisfied.  The solution $N_0(x)$ of the lapse-fixing equation\footnote{For a physical interpretation of the solution of \eqref{LFE1} as an effective ``experienced lapse" for weak matter field fluctuations see \cite{TimMatter}. For a related discussion of how spacetime emerges from coupling shape dynamics to matter see \cite{SD-Matter}.} \eqref{LFE1} is unique up to a choice of boundary conditions on the lapse. This will be discussed in further detail in chapter \ref{rindler}.

After imposing the gauge fixing condition $\pi_{\phi}(x) \equiv 0$ and working out the consistency conditions in the algebra of constraints, we are left with the first class constraints $T_{\phi}\mathcal{S}(N_0)$, $-4\pi(x)$ and $T_{\phi}\mathcal{H}_i(x) = \mathcal{H}_i(x)$ and the second class constraints $T_{\phi}\mathcal{S}(x)-T_{\phi}\mathcal{S}(N_0)\sqrt{q}(x)$ and $\pi_{\phi}(x)$. Of the remaining first class constraints, $T_{\phi}\mathcal{S}(N_0)$ generates, foliation-preserving time reparametrizations of the form $t \to f(t)$, $\mathcal{H}_i(x)$ generates diffeomorphisms acting on the conformal spatial metric and it conjugate momentum, and $-4\pi(x)$ generates spatial Weyl transformations. To summarize, the total Hamiltonian for shape dynamics is given by 
\begin{equation}\label{shapeHam}
\mathcal{H}_{\mbox{\tiny SD}} = T_{\phi}\mathcal{S}(N_0) + \mathcal{H}_i(\xi^i) -4\pi(\rho),
\end{equation}

\noindent which leads to the first-order equations of motion:
\begin{eqnarray}\label{EOM}
\dot{q}_{ij} &=& 4\rho q_{ij} + 2e^{-6\phi}\frac{N_0}{\sqrt{q}}\pi_{ij} + \mathcal{L}_{\xi}q_{ij}  \\
\dot{\pi}^{ij} &=& N_0 e^{2\phi}\sqrt{q}(R^{ij} - 2 \phi^{;ij} + 4 \phi^{;i}\phi^{;j} - \frac{1}{2}Rq^{ij} + 2\nabla^2\phi q^{ij}) \nonumber  \\
 & & -e^{2\phi}\sqrt{q}(N_0^{;ij} - 4\phi^{(,i}N_0^{,j)} - \nabla^2 N_0q^{ij}) + \mathcal{L}_{\xi}\pi^{ij} - 4\rho\pi^{ij}  \\
 & & -\frac{N_0}{\sqrt{q}}e^{-6\phi}(2\pi^{ik}\pi^{j}_{k} - \pi^{kl}\pi_{kl}q^{ij}). \nonumber \\ \nonumber
\end{eqnarray}

The non-vanishing part of the constraint algebra for shape dynamics is given by
\begin{eqnarray}
\{\mathcal{H}_i(\xi_1^i),\mathcal{H}_j(\xi_2^j)\} &=& \mathcal{H}_k([\xi_1,\xi_2]^k) \\
\{\mathcal{H}_i(\xi_1^i),\pi(\rho)\} &=& \pi(\xi^i\partial_\rho).
\end{eqnarray}

As claimed in the previous chapter, the constraint algebra for shape dynamics does not vary from point to point of phase space as it does in canonical general relativity. 

At this stage, it would be conscientious to admit some shortcomings of the formalism presented above. Since the solution of the lapse-fixing equation $N_0$ is obtained by inverting a second-order differential operator, $T_{\phi}\mathcal{S}(N_0)$ is non-local and hence $\mathcal{H}_{SD}$ is also non-local. For the same reason, it is typically a non-trivial problem even to write down the explicit form of the shape dynamics Hamiltonian for all but the simplest initial data.  

Also, while the asymptotically flat formulation of shape dynamics described above is very useful for studying isolated systems, it should be thought of as an approximation to the more fundamental, spatially compact formulation of the theory. In that spirit, the solutions studied here should be thought of as approximately describing nearly empty regions of a much larger, spatially compact universe. The primary drawback of the asymptotically flat framework is that the solutions depend on a choice of boundary conditions, which spoils some of the relationalism  that the full, spatially compact theory boasts \cite{SDanIntro}. Moreover, it is not yet clear what physical role the degeneracy introduced by the freedom to choose boundary conditions plays in the theory. The reader should particularly keep this point in mind when reading chapter \ref{rindler}.
\chapter{Static and Rotating Black Hole Solutions for Shape Dynamics}\label{BHS}

Despite the local equivalence of shape dynamics and general relativity, there nevertheless exists the possibility that corresponding solutions of shape dynamics and general relativity may have different global structures. In particular, it was recently shown in \cite{Birkhoff} that while shape dynamics possesses a unique asymptotically flat, spherically symmetric solution that agrees with general relativity near spatial infinity, the solution differs physically from the Schwarzschild solution at and within the event horizon. This chapter closely follows \cite{Kerr} in which the author, in collaboration with H. Gomes, generalized these results to the case of stationary, axisymmetric solutions, and explicitly derived the rotating black hole solution for shape dynamics. I will discuss how the solutions are related to the corresponding exterior general relativity solutions, and how they depart from one another at the horizon. Finally, I will make contact with the spherically symmetric solution by considering the zero angular momentum limit of the rotating solution.

Before considering the axisymmetric case, it is instructive to review the main results of \cite{Birkhoff} for the spherically symmetric solution of shape dynamics. 

\section{The isotropic line element}
The Birkhoff theorem for general relativity states that the only vacuum solution that is spherically symmetric is in the isometry class of the Schwarzschild line element. For shape dynamics, an analogous theorem exists, but there one must also provide boundary conditions on the dynamical variables. By reconstructing the line element one obtains: 
\begin{equation}
 \label{equ:SD_new}
ds^2= -\left(\frac{1-\frac{m}{2r}}{1+\frac{m}{2r}}\right)^2dt^2+ \left(1+\frac{m}{2r}\right)^4\left(dr^2+r^2(d\theta^2+\sin^2\theta\, d\phi^2)\right),
\end{equation}
where $\phi$ and $\theta$ are the usual angular coordinates. The radial coordinate $r$ is related to the Schwarzschild radial coordinate,  $ r_{\mbox{\tiny{s}}}$  only for $r>m/2$. In that regime the line elements are isometric, and the relation between the two radial coordinates is given by
\begin{equation}\label{Schwarz to Iso}
r_{\mbox{\tiny{s}}} = r\left(1 + \frac{m}{2r}\right)^2.
\end{equation}
At $r =m/2$, the coordinate transformation breaks down. The  line element \eqref{equ:SD_new} is degenerate at the horizon, where the 4-volume of the reconstructed metric collapses. This collapse is a coordinate-independent statement from the point of view of shape dynamics, as the vanishing of the metric determinant $\det{(g)}=0$ cannot be avoided by any combination of spatial diffeomorphisms and spatial Weyl transformations. Indeed, let M be the reconstructed spacetime associated with \eqref{equ:SD_new} and let $S = \{x\in M\,|\,r = m/2\}$. Now, suppose $\exists f\in \mbox{diff}(M)$ such that $f^{*}\bar{g} = g$ with $\det{(\bar{g})} \neq 0$ everywhere on M. But since $\det(g)=\det(df)^2\det(\bar{g}) = 0$ on $S$, and $\det(\bar{g})\neq 0$, we must have $\det(df) = 0$ on M. But then $f^{-1}$ is not differentiable on $S$ and we reach a contradiction.  This defect in the spacetime view, however, does not afflict a shape dynamics interpretation, since shape dynamics does not require the emergence of a non-singular reconstructed spacetime for a solution to be well-defined---the spatial conformal geometry (and it's conjugate momentum) are perfectly smooth.  

An observer in the asymptotically flat region will not see any difference between this solution and a Schwarzschild black hole at the classical level. Nonetheless, the fact that there is indeed a \emph{physical} difference can be seen directly by observing differences in physical statements made about general relativistic vacuum spacetime solutions. 
For instance, a simple calculation shows that the infalling radial geodesic takes infinite proper time to reach $r=0$ \cite{Poplawski}. The proper time along an ingoing radial geodesic is therefore unbounded in the interior of the shape dynamics solution in contrast to the interior of the Schwarzschild solution in general relativity in which the proper time along ingoing radial geodesics is bounded from above. It must be emphasized at this point that these arguments are somewhat heuristic. Timelike geodesics are not particularly natural objects to study in shape dynamics, and we refer to them only because of the close analogy between shape dynamics and general relativity. In fact, when viewed as a solution of general relativity, this solution has a delta function contribution to the Ricci scalar at the horizon, and as a consequence of the vanishing of $\det(g)$ the inverse metric $g^{\mu\nu}$ is ill defined on the horizon. On the other hand, the geodesic equation makes explicit use of the inverse metric, so while timelike geodesics are still well-defined across the horizon as the piece-wise smooth curves in the reconstructed spacetime maximizing the proper time, they are not well-defined as the solution of a differential equation in a neighborhood of the horizon. Despite the heuristic nature of these arguments, we will take the point of view that since the canonical variables are smooth everywhere, the solution is non-singular from the point of view of shape dynamics.

Furthermore, the line element \eqref{equ:SD_new} can be shown to form from a thin-shell collapse \cite{Tim-collapse}. This is possible because from the shape dynamics perspective it is \emph{reduced phase space}  continuity of the solutions that is fundamental, not spacetime continuity. If one demands spacetime continuity, one recovers the usual Schwarzschild solution, if one demands reduced phase space continuity of the solution, one recovers a collapse asymptoting  to the line element \eqref{equ:SD_new}. 
 
Another interesting property that will persist in the
axisymmetric case is that the transformation $r\to m^2/(4r)$  can be
checked to leave the form of the line element (\ref{equ:SD_new})
invariant. Thus the solution has the property of \emph{inversion},
associated to conformal invariance (as for instance in the method of images as applied to spherical conducting shells in electrodynamics). The presence of this inversion symmetry, along with asymptotic flatness, is enough to see that there cannot be a physical singularity at $r = 0$ in the shape dynamic solution.

It is natural to ask if this singularity-avoiding property persists in all solutions of shape dynamics. It is worth noting that spacetimes in maximal slicing have a well-known singularity avoidance property for their Eulerian observers \cite{Gourgoulhon}.\footnote{For the constant mean curvature case relevant when the spatial manifold is compact, one can still obtain cosmological, or ``crunching" singularities. } Due to the close relationship between shape dynamics and ADM in maximal slicing, there is good reason to think that shape dynamics might share this singularity-avoiding property.  One of the purposes of this paper is to show that at least in the case of axisymmetric solutions this is indeed the case. 

\section{Stationary, Axisymmetric Solutions}\label{SAS}

We will consider the stationary axisymmetric line element:
\begin{equation}\label{Axi}
ds^2 = -(N^2 - \Omega\Psi\xi^2)dt^2 + \Omega[(dx^1)^2 + (dx^2)^2 + \Psi d \phi^2] + 2\Omega\Psi\xi d \phi dt
\end{equation}
where $N$, $\xi$ and $\Omega$ are the lapse, shift and conformal factor respectively, $\Psi$ is a function that determines the entire spatial conformal geometry, and all functions depend exclusively on $x^1$ and $x^2$. Strictly speaking $\xi ^a=\xi\delta^a_{\phi}$ is the shift vector but we will from time to time abuse language and refer to the scalar $\xi$ simply as ``the shift" since it is often more convenient to work directly with this quantity.

We would like to show that the constant-$t$ hypersurfaces of the line
element \eqref{Axi} induce a maximal slicing, i.e., that the momentum conjugate to the spatial metric has vanishing trace, $\pi = 0$. 
To this end, consider Hamilton's equation for the time derivative of the spatial metric, which takes the form
\begin{equation}\label{Hamilton}
\dot{q}_{ij} = 2N(\Omega^3\Psi)^{-1/2}(\pi_{ij}-\frac{1}{2}\pi q_{ij}) + \mathcal{L}_{\xi}q_{ij}
\end{equation}     
where $\mathcal{L}_{\xi}q_{ij}$ denotes the lie derivative of the spatial metric along the shift vector. Since the spatial metric is independent of $\phi$ we have
\begin{dmath}\label{lie metric}
\mathcal{L}_{\xi}q_{ij}  =   \partial_i \xi q_{\phi j} + \partial_j \xi q_{i \phi}  =   
2\Omega\Psi{\delta^{\phi}}_{(i}\xi_{,j)} 
\end{dmath}

where we have introduced the round parentheses for symmetrization of indices and the comma for coordinate derivatives.   Putting \eqref{lie metric} into \eqref{Hamilton}, and noting that the spatial metric is independent of $t$, we obtain
\begin{equation}\label{Hamilton2}
2N(\Omega^3\Psi)^{-1/2}(\pi_{ij}-\frac{1}{2}\pi q_{ij}) + 
2 \Omega\Psi {\delta^{\phi}}_{(i}\xi_{,j)}= 0
\end{equation}

Contracting with $q^{ij}$ yields: 
\begin{eqnarray}\label{TraceHam}
2N(\Omega^3\Psi)^{-1/2}(\pi - \frac{3}{2}\pi) +
2\xi_{,i}q^{i\phi} &=& \nonumber \\ 
-N(\Omega^3\Psi)^{-1/2}\pi +2(\Omega\Psi)^{-1}\xi_{,\phi} &=& 0 
\end{eqnarray}

Noting that $\xi$ is independent of $\phi$, we find that 
\begin{equation}\label{maximal}
-N(\Omega^3\Psi)^{-1/2}\pi = 0. 
\end{equation}

Thus we find that whenever a general relativity solution can be written as \eqref{Axi}, either it is maximally sliced or $N(\Omega^3\Psi)^{-1/2}=0$. We will assume that $\Omega$ and $\Psi$ are bounded (have finite values on compact sets). If furthermore $N$ can vanish only on singular subsets of $M$, then it follows that $\pi=0$ everywhere except at most a singular subset of $M$. Continuity of $\pi$ then demands that it is zero everywhere in space. 

Furthermore, any stationary, axisymmetric solution of Einstein's equations can be put into the form \eqref{Axi} \cite{Papapetrou}, so in principle the line element can be formed by a reconstruction of a shape dynamics solution. We will see in section \ref{completeness} that indeed there are no obstructions on the shape dynamics side of the duality.  

\section{Rotating Black Hole Solution}\label{RBHS}
\subsection{The Solution}
The Kerr spacetime is a stationary, axisymmetric solution to Einstein's equations. As such, it must be possible (at least locally) to cast it  in the form \eqref{Axi}. Indeed, it was shown in \cite{Papapetrou} that the Kerr metric can be put in the form 
\begin{equation}\label{Kerr PS} 
ds^2 = -\lambda^{-1}(dt - \omega d\phi)^2 + \lambda [m^2e^{2\gamma}(d\mu^2 + d\theta^2) + s^2d\phi^2]
\end{equation}
where 
\begin{eqnarray}\label{Kerr functions PS} 
s \hspace{5pt} &=& mp\sinh\mu\sin\theta \nonumber \\
e^{2\gamma} &=& p^2\cosh^2\mu  + q^2\cos^2\theta - 1 \nonumber\\
\omega^{-1} \hspace{5pt} &=& e^{-2\gamma}\left[2mq\sin^2\theta(p\cosh\mu + 1)\right]  \\
\lambda \hspace{5pt} &=& e^{-2\gamma}\left[(p\cosh\mu + 1)^2 + q^2\cos^2\theta \right]. \nonumber \\ 
p^2 + q^2 &=& 1 \nonumber
\end{eqnarray} 
where $p = \sqrt{1-a^2/m^2}$, $q = a/m$,  $a = J/m$, and where $m$ and $J$ are the mass and angular momentum.  It is easy to see that the metric written in terms of these (prolate spheroidal) coordinates is in the form \eqref{Axi}, so it can be mapped directly onto a shape dynamics solution via ADM decomposition. The lapse and shift can be read from of the line element:
\begin{equation}\label{lapse shift}
 N^2 = \lambda^{-1}\left(\frac{\omega^2}{\lambda^2s^2-\omega^2} + 1\right), \hspace{10pt} \xi = \frac{\omega}{\lambda^2-\omega^2}.
\end{equation}
 Putting \eqref{lapse shift} into \eqref{Hamilton2} and solving for $\pi_{ij}$, we find
\begin{equation}\label{first pi}
\pi_{ij} = -\left(\frac{\Omega^5\Psi^3}{N^2}\right)^{1/2}\left[{\delta^{\phi}}_{(i}{\delta^{\mu}}_{j)}\xi_{,\mu} +{\delta^{\phi}}_{(i}{\delta^{\theta}}_{j)}\xi_{,\theta} \right].
\end{equation}
It is easy to see that at the horizon, where $s = 0$, the lapse goes
to zero. One might then worry that $\pi_{ij}$ might diverge there,
violating continuity of the phase space variables. However, we should note that $\Psi$ goes as $s^2$, so clearly the prefactor in the  \eqref{first pi} goes to zero at the horizon.

Since it is more familiar, we would like to show that \eqref{Kerr PS} is locally diffeomorphic to the Kerr metric written in Boyer-Lindquist coordinates. The coordinate transformation
\begin{equation}\label{PS to BL}
\mu = \cosh^{-1}\left(\frac{r_{\mbox{\tiny{BL}}} - m}{\sqrt{m^2 - a^2}}\right)
\end{equation}
brings \eqref{Kerr PS} into the desired form:
\begin{equation}\label{Kerr BL}
ds^2 = -\frac{\Delta}{\Sigma}\left(dt - a \sin^2\theta d\phi\right)^2 + \frac{\sin^2\theta}{\Sigma}\left((r_{\mbox{\tiny{BL}}}^2 + a^2)d\phi - a dt\right)^2 + \frac{\Sigma}{\Delta}dr_{\mbox{\tiny{BL}}}^2 + \Sigma d\theta^2
\end{equation} 
where 
\begin{eqnarray}\label{Kerr functions BL}
\Delta  &=&  r_{\mbox{\tiny{BL}}}^2 - 2mr_{\mbox{\tiny{BL}}} + a^2 \nonumber \\
\Sigma   &=& r_{\mbox{\tiny{BL}}}^2 + a^2 \cos^2\theta \\
\end{eqnarray}

 It is interesting to note that the change of coordinates \eqref{PS to BL} is purely spatial, so it would seem that the two forms are equally valid from the point of view of shape dynamics. This is not the case, however, since the transformation fails to be differentiable at the event horizon, which is conveniently labeled in prolate spheroidal coordinates by $\mu = 0$. The transformation is therefore not globally a diffeomorphism, which is compatible with the fact that the ADM decomposition of the Kerr metric in Boyer-Lindquist coordinates constitutes a shape dynamics solution only outside the event horizon. The solution written in prolate spheroidal coordinates possesses no such deficiency, and represents a complete solution of the shape dynamics equations of motion, even though it too breaks down at the event horizon when viewed from the perspective of general relativity, as we will see in following section. 

\subsection{Completeness of the Solution}\label{completeness}

It is fairly easy to see that the Kerr metric written in prolate spheroidal coordinates breaks down at the event horizon from the point of view of general relativity. To make this breakdown explicit, we need only consider the determinant of the spacetime metric.
									
\begin{equation}\label{det(g4)}
\det(g) 
= -\lambda^2m^4e^{4\gamma} s^2.
\end{equation}
Clearly, $s$ goes to zero at $\mu = 0$ while $\lambda$ and $e^{4\gamma}$ remain finite, indicating that $\det(g)$ goes to zero at $\mu = 0$ and consequently that the spacetime metric is noninvertible there. This shows that the Kerr metric written in terms of prolate spheroidal coordinates posesses a coordinate singularity at the event horizon, $\mu = 0$. This is then not a complete solution from the point of view of general relativity, but must be regarded as a solution only in the region outside of the event horizon. This is not the case, however, from the point of view of shape dynamics, where the conformal spatial geometry, rather than the spacetime geometry, is considered fundamental. It can be immediately seen that the determinant of the spatial metric is given by $\det(q) = m^4e^{4\gamma}\lambda^2(\lambda s^2-\lambda^{-1}\omega^2) \neq 0$ for all real values of $\mu$, and diverges only as $\mu \to \pm\infty$. This represents a rather dramatic departure from the general relativistic solution which requires an entirely different interior to the event horizon, possessing well known technical problems such as physical singularities and closed timelike curves. 

\subsection{Shape Dynamic Horizons: Classical Firewalls?}\label{horizons}

It is interesting to note that while the breakdown in spacetime geometry that occurs at the horizon is not forbidden by shape dynamics (in fact, it is \textit{required}), it does have some interesting consequences for infalling observers. The well known ``no drama" result of general relativity might not hold in shape dynamics because the equivalence principle is an emergent property of shape dynamics, not an axiom, and this property fails to emerge precisely at the event horizon. To better understand the nature of the horizon, consider the following argument.

Let us assume that in the interior and exterior regions, the trajectories of test observers are described by  timelike geodesics of the reconstructed line element. Since there are no outgoing timelike geodesics in the exterior region that originate in the interior region, reflection symmetry\footnote{We assume that a generic stationary shape dynamic black hole will possess inversion, or reflection symmetry about the horizon. Over-extreme black holes, which do not possess horizons are exempt from these considerations. There is some evidence to suggest that this assumption may be violated for boundary conditions other than asymptotic flatness, but as of yet no such solutions have been found. We will show in section \ref{ZAML} that the rotating shape dynamic black hole does possess reflection symmetry.} seems to demand that there are no ingoing timelike geodesics originating in the exterior region. It would appear then, that the horizon must be interpreted as the location where timelike geodesics terminate in both regions. This is not the case, however, as can be seen by noting that the lapse goes to zero at the horizon, and becomes negative in the interior region. As a result, the timelike geodesics in the interior region should be interpreted in a time-reversed fashion \cite{Poplawski}. Only in this peculiar manner can the ingoing geodesics in the exterior region  be continued to ingoing geodesics in the interior region.

The consequences of the argument presented above can be significant
for infalling observers. Rather than passing through the horizon
uneventfully, infalling observers might be able to perform a
measurement to determine the instant at which they pass into the
time-reversed parallel universe. Indeed, it can be shown that the
expansion scalar of a congruence of time-like geodesics suffers a
finite discontinuity, changing signs at the horizon
\cite{Poplawski}. Although the expansion scalar of an Eulerian congruence (i.e. one that is foliation orthogonal  in the reconstructed metric) is just the rate of change of the spatial volume element, and thus can be argued to be pure gauge, it is conceivable that for non Eulerian observers, and for instance for shear discontinuities as opposed to expansion, one would obtain a gauge-invariant statement regarding a discontinuity at the horizon.  This might provide infalling observers with a
well-defined signal of having crossed the horizon. A congruence of
infalling observers recording measurements of the volume and shape of a co-moving
ball of matter will find that the recorded values decrease up to the
horizon, at which point the co-moving ball will ``bounce" outward, in an observable manner by the congruence. 
In fact, even if one considers just the expansion scalar, since it has  a finite discontinuity at the horizon, there should be no continuous spatial diffeomorphism or conformal transformation that will remove this discontinuity. Hence, while different observers might disagree about the details of the measurement, it is possible that  all observe the volume bounce. This matter should be further investigated in the context of the infalling shell of dust. 

Since the shape dynamic description of stationary black holes requires a violation of the equivalence principle at the event horizon, it is an exciting possibility that quantum shape dynamic black holes may change the picture of the firewall paradox  \cite{AMPS}. It is too early to tell at this writing what, if any, insights shape dynamics can contribute to this debate, but the author is currently investigating the properties of shape dynamic horizons in this context.

\section{Alternative Gauge-Fixing}\label{AGF}
\subsection{Equations of Motion}
The spatial conformal invariance of shape dynamics allows us to cast the solution in an alternative form by extracting a common scalar function from the spatial metric.  The transformed metric and conformal factor are
\begin{eqnarray}\label{new gauge}
q_{ij} &=& m^2\left(\delta^{\mu}_i \delta^{\mu}_j + \delta^{\theta}_i \delta^{\theta}_j + m^{-2}e^{-2\gamma}(s^2 - \lambda^{-2}\omega^2)\delta^{\phi}_i \delta^{\phi}_j\right) \nonumber \\ 
&=&  m^2\left(\delta^{\mu}_i \delta^{\mu}_j + \delta^{\theta}_i \delta^{\theta}_j + \Psi\delta^{\phi}_i \delta^{\phi}_j\right) \\
\Omega &=& e^{-2\gamma} \lambda^{-1} = [(p\cosh\mu + 1)^2 + q^2\cos^2\theta]^{-1} \nonumber 
\end{eqnarray} 
where in the second line of \eqref{new gauge} we have defined $\Psi = m^{-2}e^{-2\gamma}(s^2 - \lambda^{-2}\omega^2)$. We know from the gauge symmetries of shape dynamics that the transformed solution must also be a solution to the shape dynamics equations of motion. For the spatially noncompact case, the equations of motion read
\begin{equation}\label{gDot}
\dot{q}_{ij} = 4\rho q_{ij} + 2e^{-6\Phi}\frac{N}{\sqrt{q}}\pi_{ij} + \mathcal{L}_{\xi}q_{ij}  \\
\end{equation}
\begin{eqnarray}\label{piDot}
\dot{\pi}^{ij} &=& N e^{2\Phi}\sqrt{q}(R^{ij} - 2 \Phi^{;ij} + 4 \Phi^{;i}\Phi^{;j} - \frac{1}{2}Rq^{ij} + 2\nabla^2\Phi q^{ij}) \nonumber  \\
 & & -e^{2\Phi}\sqrt{q}(N^{;ij} - 4\Phi^{(,i}N^{,j)} - \nabla^2 Nq^{ij}) + \mathcal{L}_{\xi}\pi^{ij} - 4\rho\pi^{ij}  \\
 & & -\frac{N}{\sqrt{q}}e^{-6\Phi}(2\pi^{ik}\pi^{j}_{k} - \pi^{kl}\pi_{kl}q^{ij}) \nonumber \\ \nonumber
\end{eqnarray}
 \noindent where $\rho$ is a lagrange multiplier associated with the conformal constraint, and $\Phi = \ln\Omega$ satisfies the Lichnerowicz-York  equation:
\begin{equation}\label{LY}
\nabla^2\Omega + \frac{R}{8}\Omega - \frac{1}{8}\pi^{ij}\pi_{ij}\Omega^{-7} = 0 
\end{equation}

We can eliminate $\rho$ by putting \eqref{new gauge} into \eqref{gDot}, taking the trace, and requiring that $\pi = 0$, which immediately yields $\rho = 0$. Putting this back into \eqref{gDot}, we find that 
\begin{eqnarray}\label{momenta}
\pi_{ij} &=& \frac{\Psi^{3/2}e^{6\Phi}}{m^2N}\left( \xi_{,\mu}\delta^{\mu}_{(i}\delta^{\phi}_{j)} + \xi_{,\theta}\delta^{\theta}_{(i}\delta^{\phi}_{j)}\right) \nonumber \\
\pi^{ij} &=& \frac{\Psi^{1/2}e^{6\Phi}}{m^2N}\left(\xi_{,\mu}\delta^{(i}_{\mu}\delta^{j)}_{\phi} + \xi_{,\theta}\delta^{(i}_{\theta}\delta^{j)}_{\phi}\right)
\end{eqnarray}

While the equations of motion are somewhat more complicated in this gauge, it can be shown that the transformed solutions \eqref{new gauge}, \eqref{momenta} do indeed satisfy \eqref{gDot}, \eqref{piDot}. One key advantage of this alternative gauge fixing is that the spatial metric now possesses only one functional degree of freedom, $\Psi$. Now the entire spatial geometry can be expressed in terms of $\Psi$ and its derivatives alone. This simplified form of the metric can be exploited for the purposes of analyzing the spatial conformal structure of the solution. In particular, it wll aid us in searching for singularities in the conformal structure.

\subsection{Conformal Regularity of the Horizon}
The simplified form of the metric arising from our change of conformal gauge produces a correspondingly simplified connection and curvature tensor. The interested reader is referred to appendix A for the calculation of these quantities. We define the Cotton tensor by
\begin{equation}\label{Cotton1}
\mathscr{C}_{ijk} := \nabla_k\left(R_{ij} - \frac{1}{4}R\,q_{ij}\right) - \nabla_j\left(R_{ik} - \frac{1}{4}R\,q_{ik}\right)
\end{equation}

\noindent where $\nabla$ denotes covariant differentiation with respect to the spatial metric. The rank-two Cotton-York tensor, $C^{ij}$, can be defined by its relation to the Cotton tensor:
\begin{equation}\label{Cotton}
C^{ij} := -\frac{1}{2}q^{mj}\epsilon^{ikl}\mathscr{C}_{mkl}
\end{equation}

The Cotton tensor contains all of the local information on the
conformal geometry of a three-dimensional Riemannian manifold
\cite{Gourgoulhon} in much the same way that the Weyl tensor (which
vanishes identically in three dimensions) captures local information on the conformal geometry in higher dimensions. Like the Weyl tensor in higher dimensions, the Cotton tensor is completely traceless, conformally invariant, and vanishes if and only if the manifold is conformally flat.  From \eqref{Cotton-York}, \eqref{CY squared}, and \eqref{Cotton} we see that if $C^2 := C^{ij}C_{ij}$ diverges then there must be a singularity in the Cotton tensor. A singularity in the Cotton tensor would signal the presence of a breakdown of the conformal geometry, i.e it would be a \textit{physical} singularity from the perspective of shape dynamics. It is therefore useful to show that $C^2$ is finite as a heuristic argument that no such physical singularities are present. In this sense, although $C^2$ is not strictly speaking conformally invariant (it transforms as $C^2 \to \Omega^{5/2}C^2$  under $q_{ij} \to \Omega q_{ij}$), it can be thought of in analogy with the Kretschmann invariant in general relativity. Moreover, if we assume that $\Omega$ is bounded in the sense described in section \ref{SAS}, then the presence of conformal covariance as opposed to conformal invariance is essentially irrelevant for the purposes of identifying singularities in the conformal structure.  

Intuition suggests that the points we should scrutinize most carefully are the horizon and the limit as $\mu \to -\infty$, since we have already noted some peculiaraties about the former, and the latter seems analogous to the singularity in general relativity. We should note, however, that the latter is isomorphic to spatial infinity, so asymptotic flatness ensures that there are no conformal singularities there. 

 To help simplify the calculation for the horizon, we can note that since $\Psi$ is an even function of $\mu$, any odd number of $\mu$ derivatives acting on $\Psi$ will be zero when evaluated on the horizon. Taking this into account we can put \eqref{CY squared} in the simplified form
\begin{equation}\label{CY squared on horizon}
C^2(0, \theta) =  \left.\left[ \frac{1}{4\Psi^2}\Psi_{,\theta}\Psi_{,\theta\theta} - \frac{1}{4\Psi}\left( \Psi_{,\mu\mu\theta} + \Psi_{,\theta\theta\theta} \right) \right] \right|_{\mu = 0}
\end{equation}

\noindent At the horizon, we have 
\begin{equation}\label{psi on horizon}
\Psi(0, \theta) = \frac{4a^2\sin\theta}{m^2\left(4 + q^2\cos^2\theta\right)^2}
\end{equation}

\noindent which is nonzero except on the axis of rotation $\theta = \{0, \pi\}$. We will therefore have to carefully analyze the limits as we approach these points. A lengthy but straightforward calculation yields the other ingredients of \eqref{CY squared on horizon}:
\begin{align}\label{ingredients}
\Psi_{,\theta}(0,\theta) &= -\frac{2^8 a^2 q^2 \sin^2\theta\cos\theta}{m^2(4+q^2\cos^2\theta)^3} \nonumber \\
\Psi_{,\theta\theta}(0,\theta) &= \frac{2^8}{m^2(4+q^2\cos^2\theta)^3}\left[ a^2q^2\sin^2\theta(\sin^2\theta-\cos^2\theta) - \frac{6a^2q^4\sin^3\theta\cos^2\theta}{4+q^2\cos^2\theta} \right] \nonumber \\
\Psi_{,\theta\theta\theta}(0,\theta) &=  \frac{2^9a^2q^2\sin^2\theta\cos\theta}{m^2(4+q^2\cos^2\theta)^4}\Big[ 3\cdot2^3q^4\sin^2\theta\cos^2\theta + \nonumber \\ & \hspace{45mm} 3^2q^2(\sin^2\theta-\cos^2\theta) + 2(4+q^2\cos^2\theta) \Big] \nonumber \\
\Psi_{,\theta\mu\mu}(0,\theta) &= 4\sin\theta\cos\theta + 3\cdot2^{10}\frac{a^2q^2}{m^2}\frac{\sin\theta\cos\theta}{(4+q^2\cos^2\theta)^4}
\end{align}

Clearly, none of the functions in \eqref{ingredients} can diverge for any values of $\theta$. Moreover, if we insert \eqref{ingredients} back into \eqref{CY squared on horizon}, we see that $C^2 = 0$ at $\theta = 0$ and $\theta = \pi$. So despite the peculiar behavior of the horizon when viewed from the four-dimensional perspective, we do not see any conformal singularities manifesting themselves in the Cotton-York tensor at the horizon.  

\section{Zero Angular Momentum Limit}\label{ZAML}

Finally, it will be demonstrated that in the zero angular momentum limit $a = 0$ of the solution presented in section \ref{RBHS}, we recover the spherically symmetric solution \eqref{equ:SD_new} presented in \cite{Birkhoff}. From the definitions of $p$ and $q$, we can see that $a = 0$ implies $p = 1$, $q = 0$. Putting these limits into \eqref{Kerr functions PS} we obtain
\begin{align}\label{PS a = 0}
&s \hspace{7.7pt}= \hspace{7pt} m\sinh\mu\sin\theta \nonumber \\
&\lambda \hspace{7pt} = \hspace{7pt} \frac{(\cosh\mu + 1)^2}{\sinh^2\mu}  \\
&\omega \hspace{7pt} = \hspace{7pt} 0 \nonumber \\
&e^{2\gamma} = \hspace{7pt} \sinh^2\mu \nonumber.
\end{align}  

Inserting \eqref{PS a = 0} into \eqref{Kerr PS} gives the spherically symmetric line element written in terms of prolate spheroidal coordinates.
\begin{equation}\label{Kerr PS a = 0}
ds^2 =  -\frac{\cosh^2\mu + 1}{(\cosh\mu + 1)^2}dt^2 +  m^2(\cosh\mu + 1)^2\left( d\mu^2 + d\theta^2 + \sin^2\theta d\phi^2\right).
\end{equation}

One can already see the isotropic character of the solution as written
in terms of prolate spheroidal coordinates. In order to obtain the
form of the solution presented in \cite{Birkhoff} we perform the
spatial diffeomorphism $r = \frac{m}{2}e^{\mu}$ which can be checked
to reproduce \eqref{equ:SD_new}, as desired. It is important to note
that unlike \eqref{PS to BL}, this transformation and its inverse are
differentiable  everwhere, so it is a global diffeomorphism-- i.e. the transformation is pure gauge. It is interesting to note that the inversion symmetry of the spherically symmetric solution takes on a simplified form when written in terms of prolate spheroidal coordinates. In this case the inversion symmetry is manifested by the fact that the line element is an even function of $\mu$. Indeed, this is the case even before we take the limit $a = 0$, so we conclude that the axisymmetric solution also possesses an inversion symmetry under $\mu \to -\mu$. This is a nice representation of the symmetry since it emphasizes that what we are doing is \emph{reflecting} about the event horizon $\mu = 0$ into the corresponding point of the time-reversed mirror universe.

\section{Discussion}
The most general local form of stationary, axisymmetric vacuum solutions to the shape dynamics equations of motion have been derived and this result was used to obtain the rotating black hole solution for shape dynamics. The rotating black hole solution preserves many of the striking features of the spherically symmetric case. It possesses a powerful inversion symmetry about the horizon where it does not form a spacetime, and it seems to completely avoid physical singularities. The inversion symmetry and singularity avoidance are perhaps even more surprising in the rotating case, since the corresponding general relativity solution is so complicated in the interior region, possessing a ringlike physical singularity, closed timelike curves and an inner Cauchy horizon. The shape dynamics solution, by contrast, avoids all of these difficulties by creating at the event horizon the time-reversed mirror universe that allows the matter source to expand out to an inner spatial infinity and avoid collapsing to a singularity. 

The extreme and over-extreme Kerr solutions can be mapped onto their
shape dynamic counterparts using the same arguments presented
above. These solutions are also presented in \cite{Papapetrou} in the
form \eqref{Axi}, making the mapping to shape dynamics almost
trivial. In the overextreme case, the four dimensional line element
associated with the shape dynamics solution is globally related to
Boyer-Lindquist coordinates by a spatial diffeomorphism. It is not yet
clear whether the naked singularity persists in the shape
solutions---it is likely that this singularity is fundamentally four
dimensional in nature and does not appear in shape dynamics.

Probably the most exciting feature of the black hole solutions for shape dynamics is that they do not form a spacetime at the horizon. Susskind and Maldacena have argued \cite{Cool} that there is close relationship between entanglement and (non-traversable) wormholes in the context of a possible resolution to the firewall paradox. In shape dynamics, the stationary black hole solutions seem to be traversable wormholes\footnote{A more definitive answer to the question of traversability can only be answered by coupling matter degrees of freedom. This has been done in the spherically symmetric case for a collapsing thin shell of matter \cite{Tim-collapse}. } that suggest two sources for a possible resolution to this debate: black holes are correctly described by shape dynamics, and there is no paradox because the equivalence principle breaks down at the horizon and/or there is no singularity. The obvious next steps in these considerations are to look at the semiclassical behavior of these solutions, to analyze their thermodynamic properties, and to consider the behavior of quantum fields in the presence of a stationary shape dynamic black hole background. 

The fact that the solution is well-behaved at the horizon gives it an advantage over similar traversable wormhole models in general relativity, which are generically unstable. Furthermore, since the latter require a delta function contribution to the curvature scalar at the horizon \cite{Visser}, these solutions must be regarded as singular spacetimes. Moreover, if traversable wormholes admit a more consistent quantum mechanical interpretation than the standard stationary black hole solutions in general relativity, then since these solutions arise naturally in shape dynamics\footnote{It should be noted that these seem to be the \emph{only} stationary black hole solutions that arise naturally in shape dynamics. The singularity avoidance theorems for Eulerian observers in maximal slicing make it implausible that anything like ordinary Schwarzschild or Kerr solutions could be made to satisfy the shape dynamics equations of motion everywhere.} this might be a hint that shape dynamics is a more consistent classical theory of gravity than general relativity for the purposes of quantization. As a last remark, let us mention that since the shape dynamics  solutions are indistinguishable from the corresponding general relativity solutions in the asymptotically flat region, they are on equally solid ground from the point of view of current empirical astrophysical observations.
\chapter{Towards Shape Dynamic Black Hole Entropy}
It has been shown that the simplest asymptotically flat black hole solutions admitted by shape dynamics disagree with general relativity at and within their event horizons, and that a smooth spacetime geometry fails to emerge on the horizon. Moreover, the areas of the horizons in the shape dynamic black hole solutions are not invariant under spatial Weyl transformations. The fact that these areas are not gauge-invariant quanitities has cast considerable doubt on the prospect of recovering the famous result that the entropy of a black hole is equal to one quarter of the horizon area. The  purpose of this chapter, which is closely based on a paper by the author and V. Shyam \cite{Vasu} is to show that by carefully treating the horizon as an interior spatial boundary and analyzing the boundary terms needed to obtain well-defined equations of motion in the sense first described by Regge and Teitelboim \cite{Regge}, one can indeed recover the result that $S=A/4$ for a shape dynamic black hole. 

This is an important result for shape dynamics in light of the fact that the proportionality of black hole entropy and horizon area has been derived independently by state-counting arguments in a diverse array of approaches to quantum gravity each of which count very different states to arrive at the same result. The fact that these very different approaches universally yield the same result is something of an interesting problem in and of itself, and it has been argued by Carlip \cite{Carlip, universality} and others that this ``problem of universality" may arise as a result of a near-horizon conformal symmetry in the Einstein-Hilbert action. At any rate, the ubiquity of the thermodynamic properties of black holes has been so thoroughly demonstrated at the theoretical level, that while we have yet to detect Hawking radiation by direct observation, one would rightly be deeply skeptical of any theory of gravity in which black holes do not admit this standard thermodynamic interpretation.

\section{Boundary Hamiltonian, Euclidean Gravity and Black Hole Entropy}
It was first pointed out by Regge and Teitelboim \cite{Regge} that in order to properly formulate canonical general relativity with spatially non-compact boundary conditions, one must add a boundary term $H_B$ to the pure constraint ADM Hamiltonian $H_0$ so that the total Hamiltonian is given by the sum 
\begin{equation}
H = H_0 + H_B.
\end{equation}

\noindent If one neglects the boundary term $H_B$ no well-defined equations of motion are generated by $H_0$---only the total Hamiltonian $H  = H_0 + H_B$ yields well-defined equations of motion. This is essentially due to the fact that when calculating Poisson brackets on a manifold with boundary, one cannot discard the boundary terms that arise due to integration by parts. Rather, these terms must be canceled with appropriate boundary terms so that Hamilton's equations can be recovered. Moreover, given mild restrictions on the fall-off of the surface deformations, the boundary term $H_B$ can be identified with the total energy of the system. Let $\Sigma$ be a constant-$t$ hypersurface with boundary $\partial\Sigma$. Including all boundary terms, the ADM action can be written as
\begin{eqnarray}\label{ADM action}
\mathcal{I}_{\mbox{\tiny ADM}} &=& \frac{1}{16\pi}\int dt \int_{\Sigma} d^3x\sqrt{q} \left(\dot{q}_{ij}\pi^{ij} -
  N\mathcal{S}(x) - \xi^i\mathcal{H}_i(x)\right) \nonumber \\ &-&
\frac{1}{8\pi}\int dt\int_{\partial\Sigma} d^2x\sqrt{\sigma}(NK+r^iN_{,i}-r_i\xi^j\pi^i_j)
\end{eqnarray}

\noindent where $\sigma_{ij}$ (which is held fixed, i.e. $\delta \sigma_{ij} = 0$) is the metric induced on $\partial\Sigma$ by $q_{ij}$, $r^i$ is the outward pointing normal vector of  $\partial\Sigma$, $K$ is the trace of the extrinsic curvature of $\partial\Sigma$ embedded in $\Sigma$, and $N$ and $\xi^i$ are the lapse function and shift vector.

The connection between gravity and thermodynamics was first proposed by Bekenstein in 1972 \cite{Bekenstein} and confirmed by Hawking in 1974 through the discovery of thermal black hole radiation (Hawking radiation) \cite{HawkingRadiation}. In 1976, Gibbons and Hawking introduced the  Euclidean canonical formalism in \cite{Hawk}. In this work, it was shown that the path integral associated with the Euclideanized gravitational action has the exact form of a canonical partition function at inverse temperature $\beta$ where the path integral is taken over all field configurations that are periodic with period $i\beta$. From the lowest order contributions to the canonical partition function, one can then obtain all of the thermodynamic quantities associated with the horizons of classical solutions to the Einstein field equations, including the entropy.  

Since these seminal works, many other approaches to black hole thermodynamics have been developed. Here, I will make use of an approach developed by Padmanabhan \cite{HoloPad} that identifies the boundary action (or Hamiltonian) evaluated on the horizon of a stationary black hole with the entropy (or the temperature times the entropy) of the black hole. In this approach, one considers the fact that horizons generically appear for some families of observers (such as accelerated observers in flat spacetime, and Eulerian observers in the exterior of the Schwarzschild spacetime) but not for others (such as inertial observers in flat spacetime or free-falling observers in the Schwarzschild spacetime). One then argues that any family of observers should be able to use an action principle that makes use only of the information available to those observers. Since observers who see a horizon cannot receive information from the other side of the horizon, the action integral for observers who see horizons should only depend on the portion of the spacetime outside the horizon. Since there is a ``tracing out" of information about what is on the other side of the horizon when one uses such a family of observers, it is natural to identify the boundary term in the action coming from the horizon with an entropy, and one indeed finds that for a stationary horizon with $N = 0$, $N_{,i} =  \kappa r_i$ and $\pi^{ij} = 0$ where $\kappa$ is the surface gravity of the horizon, the boundary action equals one quarter of the horizon area.  

Now we would like to show that the same argument can be used to relate the horizon contribution to the shape dynamics Hamiltonian with the entropy of a shape dynamic black hole. The boundary Hamiltonian for shape dynamics was derived in \cite{poincare} and its variation was found to be 
\begin{multline}\delta t_\phi H_B(N,\xi)= \\ 2\int_{\partial\Sigma} d^2 y\, \xi^i r^j\left\{\pi^{kl}\left((5q_{ik}q_{jl}-q_{ij}q_{kl})\delta\phi+(q_{jl}\delta q_{ki}-\frac{1}{2}q_{ij}\delta q_{kl})\right)+q_{ik}q_{jl}\delta\pi^{kl}\right\}\\
+\int_{\partial\Sigma} d^2 y\sqrt \sigma e^{2\phi}\Big\{8\left( N\delta\phi^{,i}-N^{,i}\delta\phi\right)r_ a  \\ + \left( N\delta q_{ij;l}+(6\phi_{,l} N-N_{,l})\delta q_{ij}\right)(q^{lm}q^{ij}-q^{li}q^{jm}) r_m \Big\} \label{Boundary Variation 1}
\end{multline}

\noindent  Eulerian observers\footnote{ An Eulerian observer, also known as a hypersurface orthogonal observer or zero angular momentum observer (ZAMO), is an observer whose four-velocity $u^{\mu}$ is perpendicular to $\Sigma_t$, i.e $u^{\mu}\epsilon_{\mu\nu\rho} = 0$, where $\epsilon_{\mu\nu\rho}$ is the volume form on $\Sigma_t$.} are natural observers to choose in shape dynamics because they are at rest with respect to the spatial geometry. These observers see a horizon in stationary black hole solutions, so it is appropriate to include a boundary contribution to the Hamiltonian from the horizon. For Eulerian observers, one can use the same boundary conditions used in general relativity:  $N = 0$, $N_{,i} =  \kappa r_i$ and $\pi^{ij} = 0$. When these boundary conditions are imposed, the total variation of the boundary Hamiltonian becomes:
\begin{gather}
\delta t_{\phi}H_B(N,\xi) = -\frac{\kappa}{16\pi}\int_{\partial\Sigma}d^2
y\sqrt{\sigma}e^{2\phi}\left( 8\delta\phi +
  r_l r_m\delta q_{ij}(q^{lm}q^{ij}- q^{li}q^{jm})  \right) \nonumber \\
=  -\frac{\kappa}{16\pi}\int_{\partial\Sigma}d^2
y\sqrt{\sigma}e^{2\phi}\left( 8\delta\phi + (q^{ij} - r^i r^j)\delta
  q_{ij} \right) \label{boundary variation 2} \\
= -\frac{\kappa}{16\pi}\int_{\partial\Sigma}d^2
y\sqrt{\sigma}e^{2\phi}\left(8\delta\phi + \sigma^{ij}\delta
  q_{ij} \right). \nonumber \\ \nonumber
\end{gather}

\noindent On the other hand, 
\begin{eqnarray}\label{detVar}
\delta\sqrt{\sigma} &=& \frac{1}{2}\sqrt{\sigma}\sigma^{ij}\delta\sigma_{ij} =
\frac{1}{2}\sqrt{\sigma}\sigma^{ij}\delta\left( q_{ij} - r_i r_j 
\right) \nonumber \\
&=& \frac{1}{2}\sqrt{\sigma}\sigma^{ij}\left(\delta q_{ij} -
  r_i r^k\delta q_{jk} - r_j r^k\delta q_{ik} \right) \nonumber \\
&=&  \frac{1}{2}\sqrt{\sigma}\sigma^{ij}\delta q_{ij}.
\end{eqnarray}

\noindent Equation \eqref{detVar} shows that the factor of $8$ appearing in the first term of the last line of \eqref{boundary variation 2} should actually be a $4$ in order for  \eqref{boundary variation 2} to be a total variation. It is absolutely essential that  \eqref{boundary variation 2} be a total variation in order for the boundary conditions we imposed to be consistent, since this term is \emph{by definition} the variation of the boundary contribution to the Hamiltonian. This suggests that we need to impose additional boundary conditions on the conformal factor $\phi$ at the horizon. An obvious choice is $\phi = 0$, $\delta \phi = 0$ since then we would fully break Weyl invariance on the
horizon, and our horizon Hamiltonian would then be simply
\begin{equation}\label{TS}
H_{hor}(N) = \frac{\kappa}{8\pi}\int_{\partial\Sigma}d^2y\sqrt{\sigma} = \frac{\kappa}{2\pi}\cdot\frac{A}{4}
\end{equation}

\noindent where $A$ is the same horizon area we would have found in general relativity. Now, if we identify $T =  \frac{\kappa}{2\pi}$, $S = \frac{A}{4}$, we find complete agreement with the horizon thermodynamics of general relativity. 

One might worry, however,
that this condition is too restrictive. We could consider restricting
$\delta\phi$ by requiring that it depend on $\delta \sigma_{ij}$. For
example, if we demand that $\delta\phi = \lambda\sigma^{ij}\delta\sigma_{ij}$ for some constant $\lambda$. However, this
approach would require the conformal factor to transform by a
$\lambda$-dependent rescaling on the horizon, which can be shown to be inconsistent with our boundary conditions at the level of the
constraints. The source of this inconsistency comes from the fact that
the bracket of the horizon Hamiltonian with the conformal constraint
can vanish only if $\langle\rho\rangle_{\sigma} :=
\frac{\int_{\partial\Sigma}d^2y\sqrt{\sigma}\rho}{\int_{\partial\Sigma}d^2y\sqrt{\sigma}}
= 0$, which implies that the spatial Weyl transformations which are pure
gauge in the presence of a horizon with the boundary conditions we
have specified are exactly those that preserve the area of the
horizon. 

In anticipation of this restriction on the Weyl invariance of shape
dynamics in the presence of a horizon, we can obtain the same outcome
as if we had set $\delta\phi = 0$ without having to make quite such a
rigid restriction. Suppose we require that
\begin{equation}
\int_{\partial\Sigma}d^2y\sqrt{\sigma}\delta\left(e^{2\phi}\right)
= 0.
\end{equation}

\noindent If we further demand that $\phi = 0$ lies in our solution space on the
horizon, so that we may recover a non-trivial intersection with GR,
then it is clear that what we are really demanding is not that $\phi$ be fixed on the horizon, but rather that the allowed
transformations of $\phi$ are those that preserve the horizon area. With these additional restrictions in place, we can integrate
\eqref{boundary variation 2} to obtain
\begin{equation}
H_{hor}(N) = \frac{\kappa}{8\pi}\int_{\partial\Sigma}d^2y\sqrt{\sigma}e^{2\hat{\phi}}
\end{equation}

\noindent where $\hat{\phi} :=
\frac{1}{2}\ln\langle e^{2\phi}\rangle_{\sigma}$, $\phi$ is an
arbitrary conformal factor satisfying the usual asymptotically flat
boundary conditions at spatial infinity, and $\langle \cdot \rangle_{\sigma}$ denotes the average over the horizon with respect to the induced metric $\sigma$. Due to the area-preserving
nature of the allowed Weyl transformations, the horizon Hamiltonian
can be written with or without the smearing $e^{2\hat{\phi}}$:
\begin{equation}
H_{hor}(N) = \frac{\kappa}{8\pi}\int_{\partial\Sigma}d^2y\sqrt{\sigma}
\end{equation}

\noindent which is identical the corresponding result for general
relativity. However, here we had to impose boundary conditions in two
stages. First, we imposed boundary conditions on the momentum, and on
the lapse and its normal derivative, in order to capture the
essential features of a horizon. Next, we were forced to impose
boundary conditions on the conformal factor as a consistency condition for
the functional integrability of the total boundary variation
\eqref{Boundary Variation 1}. 

Next, we will show that the horizon Hamiltonian we have
derived puts restrictions on the smearing $\rho$ of the conformal
constraint $\pi(\rho)$, by calculating $\{H_{hor}(N), \pi(\rho)\}$
and observing that it only vanishes for a certain class of smearings
$\rho$. We find (see appendix B) that
\begin{eqnarray}\label{total bracket}
\{H_{hor}(N), \pi(\rho)\} &=& -\frac{3}{16\pi}\int_{\Sigma}d^3x\,
\frac{\rho N}{\sqrt{q}}e^{-6\hat{\phi}} G_{ijkl}\pi^{ij}\pi^{kl} \nonumber \\
&=&
-\frac{3}{2}H_{hor}(\rho N).
\end{eqnarray}

Since the horizon Hamiltonian is weakly non-vanishing, \eqref{total bracket} tells us that the bracket between the horizon Hamiltonian and the conformal constraint cannot weakly vanish for arbitrary smearings $\rho$. It is possible to derive an elliptic differential equation for $\rho$ so that \eqref{total bracket} vanishes, but this characterization of the allowed spatial Weyl transformations is not particularly illuminating and technically difficult to analyze. However, it is possible to ``softly" modify the constraints of theory in such a way that we can clearly show that in the presence of a horizon satsfying the boundary conditions above, the spatial Weyl transformations that are pure gauge are precisely those that preserve the horizon area. This will be the subject of the following section.

\section{Modified Constraints and Area-Preserving Weyl Transformations}
Another way to obtain a gauge-invariant horizon entropy for shape dynamics is to softly modify the the conformal constraint in such a way that the modification repects the boundary conditions and leaves the horizon area invariant. The simplest realization of these conditions is in writing the conformal constraint in the following manner:
\begin{equation}\label{modConstr}\pi(\rho)=\int_{\Sigma}d^{3}x\, \rho q_{ij}\pi^{ij} -\int_{B}d^{2}y\,\rho \langle q_{ij}\pi^{ij} \rangle_{\sigma}.\end{equation}
Here we have just rewritten the usual conformal constraint with the addition of a new term which is identically zero on the interior boundary due to the boundary conditions used to characterize the horizon, i.e. given that $\pi^{ij}|_{B}=0$ the term within the angular brackets is strongly equal to 0. Note that $\frac{\delta \pi(\rho)}{\delta \rho}=0=q_{ij}\pi^{ij}$ keeping in mind the boundary condtitions imposed, so we are ensured that the constraint is not altered in any violent manner. It is convenient to introduce the weakly non-degenerate symplectic structure on the shape dynamics phase space to ease our computations. The symplectic structure for shape dynamics is defined by the two-form
\begin{equation}\Omega_{SD}=\int_{\Sigma} \delta \pi^{ij}\wedge \delta q_{ij}.\end{equation}
The Hamiltonian vector field $X_{f}$ correponding to an arbitrary phase space function $f[q_{ij},\pi^{ij}]$ is defined through the equation
$$\Omega_{SD}(X_{f})=\delta f.$$
The integral curves of the Hamiltonian vector field correpond to its flow on phase space, and the vector field is identified with the generator of infinitesimal transformations along said flow. When this vector field corresponds to a first class constraint, it generates gauge transformations along the orbits of this constraint on phase space. The Hamiltonian vector field for
the conformal transformations is thus given by 
\begin{equation}X_{\pi(\rho)}=(\rho-\langle \rho\rangle_{\sigma})q_{ij}\frac{\delta}{\delta q_{ij}}+(\rho-\langle \rho\rangle_{\sigma})\pi^{ij}\frac{\delta}{\delta \pi^{ij}},\end{equation}
which satisifies the equation $$\Omega_{SD}(X_{\pi(\rho)})=\delta \pi(\rho).$$
Thus, the object given by \eqref{modConstr} generates infinitesimal spatial Weyl transformations on the shape dynamics phase space. A simple calculation yields
\begin{equation}\label{apres}X_{\pi(\rho)}A=0\end{equation}
where $A$ is the area of the horizon. This unambiguously shows that if the right form of the conformal constraint is chosen keeping in mind the boundary conditions, the seemingly ``soft" modification of the constraint produces the correct flow on phase space. Given that the Weyl transformations generated by the constraint given by \eqref{modConstr} preserves the horizon area, in terms of large spatial Weyl transformations, \eqref{apres} is equivalent to the statement
$$\int d^{2}y\, e^{2\phi}\sqrt{\sigma} =\int d^{2}y\,\sqrt{\sigma} =A.$$
This implies for the variation of the conformal factor at the boundary that:
\begin{equation}\int_{B}d^2y \,\delta \phi\sqrt{\sigma}=0.\end{equation}
Going back to \eqref{boundary variation 2}, we find that now, 
$$\delta t_{\phi}H_B(N,\xi)=-\frac{\kappa}{8\pi}\int_{\partial \Sigma}d^{2}y \sqrt{\sigma} \sigma^{ij}\delta q_{ij},$$
as in general relativity, and from this, we can obtain the surface Hamiltonian for the horizon 
$H_{hor}=TS=\frac{\kappa A}{8\pi}.$
Identifying the Hawking--Unruh temperature $T=\frac{\kappa}{2\pi},$ the entropy is:
\begin{equation}S=\frac{A}{4},\end{equation}
which is completely identical to the area--entropy law of General Relativity, just as we derived by more direct but less elegent means in the preceding section.

\section{Discussion}
We have shown that by correctly identifying the spatial Weyl transformations that remain pure gauge in the presence of a horizon which acts as an interior spatial boundary, it is possible to identify the gauge-invariant quantity that we have suggestively labeled $S$ and which we believe plays the role of a thermodynamic entropy associated with a shape dynamic black hole. Furthermore, if we assume that that this quantity really is the entropy associated with a shape dynamic black hole, then we can reproduce the standard area-entropy relation of general relativity and various approaches to quantum gravity. It is worth stressing that while our $S$ seems to have many features that suggest it can be interpreted as an entropy (it arises from ``tracing out" information about the interior that exterior observers do not have access to, it agrees with standard results from black hole thermodynamics, it is gauge-invariant, etc.) I must admit that until it can be shown that shape dynamic black holes emit Hawking radiation and that the temperture and entropy agree with the $T$ and $S$ we have identified, there is insufficient evidence to conclude that these quantities really play the role of a physical temperature and entropy. 


It was conjectured in \cite{Kerr} that since the equivalence principle is an emergent property of shape dynamics, rather than an axiom as in general relativity, and that this property fails to emerge precisely on the event horizon of a shape dynamic black hole, that a quantum theory of gravity based on a canonical quantization of shape dynamics might present a possible resolution to the firewall paradox introduced by Almheiri et. al. \cite{AMPS}. In order to address this question in a more systematic manner, it is once again essential to understand the thermodynamic properties of shape dynamic black holes including the radiative properties of quantum fields in a shape dynamic black hole background (i.e. Hawking radiation).

The next step in the investigation of the thermodynamic properties of shape dynamic black holes is to determine whether shape dynamic black holes produce Hawking radiation. A preliminary analysis of the features of shape dynamic black holes and the mechanism by which black holes produce Hawking radiation in general relativity makes it seem very likely that shape dynamic black holes produce thermal Hawking radiation in essentially the same way as their general relativisic counterparts. Ordinary Hawking radiation can be realized by coupling a scalar field to the Einstein-Hilbert action and analyzing the resulting covariant wave equation on the black hole background. One then identifies the zeroes of the Regge-Wheeler potential where the covariant wave equation reduces to the usual Klein-Gordon equation in flat spacetime, and expands the solutions in Fourier modes. Since the zeroes of the Regge-Wheeler potential occur at spatial infinity and on the event horizon of a Schwarzschild black hole, and since the exterior region of a Schwarzschild black hole is isomorphic to the exterior of the spherically symmetric black hole solution derived in \cite{Birkhoff}, it seems very likely that standard Hawking radiation can be recovered for shape dynamic black holes. A more careful and systematic treatment of this problem is left for future work.
\chapter{Parity Horizons and Charged Black Holes}
Originally \cite{GGK}, shape dynamics was viewed as a reformulation of canonical general relativity that traded the (on-shell) refoliation invariance generated by the quadratic Hamiltonian constraint for spatial Weyl invariance generated by a new, linear Weyl constraint. The theories were quickly seen to agree for a broad class of known solutions to general relativity---particularly those admitting globally defined foliations by spatial hypersurfaces of constant mean extrinsic curvature.\footnote{For compact spatial manifolds, shape dynamics agrees with canonical general relativity when spacetime is foliated by hypersurface surfaces of constant mean extrinsic curvature. For asymptotically flat spatial manifolds, such as those considered here, shape dynamics agrees with general relativity when spacetime is foliated by maximal slices---meaning that the trace of the momentum conjugate to spatial metric vanishes.} However, questions remained about what the corresponding solutions of shape dynamics would look like for spacetimes that admitted such foliations only locally, or whether it was even possible for shape dynamics to describe such systems.

A partial answer to this question was provided by \cite{Birkhoff} and \cite{Kerr}, in which spherically symmetric and rotating asymptotically flat black hole solutions of shape dynamics were presented. These novel solutions were shown to agree with their general relativistic counterparts outside their event horizons, but behave differently in their interior regions. Both solutions possess an inversion symmetry about their event horizons, which implies that the interior regions should be interpreted as a time-reversed copy of the exterior regions, endowing the solutions with the character of a \emph{wormhole}. I will show in the remaining chapters, based on my paper \cite{Parity}, that these features are not peculiar to event horizons, but arise naturally in solutions of shape dynamics with other types of horizons as well. I analyze the properties of these horizons and show that all of them are parity horizons, which are defined in \ref{SDBH}. 

One way to understand the disagreement between shape dynamics and general relativity when solutions develop horizons is to note that in the canonical formalism, many types of horizons arise when the lapse function either vanishes or diverges. Since the determinant of the spacetime metric can be written $\sqrt{-g} = N\sqrt{q}$ where $q$ is the determinant of the spatial metric and $N$ is the lapse function, it is clear that on a horizon where the lapse vanishes or diverges, the determinants of the spatial and spacetime metrics cannot simultaneously be finite and non-zero---one must choose between a smooth spacetime and a smooth conformal spatial geometry. 

With the benefit of hindsight, it is not particularly surprising that shape dynamics and general relativity disagree when solutions of either develop horizons, however it is still interesting to consider the particular ways in which the theories disagree in the presence of such surfaces.

In the following section, I will review some properties of the asymptotically flat black hole solutions for shape dynamics that have been introduced so far, and discuss how they motivate the definition of ``parity horizons." In \ref{charge}, I introduce a spherically symmetric, charged black hole solution for shape dynamics analogous to the Reissner-N\"{o}rdstrom black hole of general relativity, and discuss the properties of the solution. In chapter \ref{rindler}, I discuss the Rindler chart over Minkowski spacetime and introduce a new solution of shape dynamics which shares many of the same features. I will discuss how this new solution differs from its general relativistic counterpart and emphasize the crucial role played by the boundary conditions in shape dynamics. In chapter \ref{bonner}, I briefly review closed timelike curves and discuss a family of solutions to Einstein's equations known as the Van Stockum-Bonner spacetimes, which except for the lowest order case, are stationary, asymptotically flat and possess a compact\footnote{Here, ``compact" means that the Cauchy horizon is compact when viewed as a (degenerate) two-surface embedded in any of the preferred spatial slices.}  Cauchy horizon within which closed timelike curves pass through every point. I will then present a novel solution of shape dynamics which agrees with the next-to-lowest order Bonner spacetime outside the horizon, but which does not develop closed timelike curves within---the Cauchy horizon is replaced by a parity horizon in the shape dynamics solution. I argue that this is evidence of a general chronology protection mechanism in shape dynamics that is significantly more parsimonious than many of the arguments that have been made for chronology protection in general relativity.

\section{Parity Horizons}\label{SDBH}
Recall from chapter \ref{BHS} that if we assume asymptotically flat and Lorentz-invariant boundary conditions,\footnote{For a more general analysis of spherically symmetric, asymptotically \emph{spatially} flat solutions of shape dynamics that are not necessarily asymptotically Lorentz-invariant see \cite{FateOfBirkhoff}.} shape dynamics admits a unique spherically symmetric solution that can be described globally by the Schwarzschild line element in isotropic coordinates:
\begin{equation}\label{isotropic}
ds^2= -\left(\frac{1-\frac{m}{2r}}{1+\frac{m}{2r}}\right)^2dt^2+ \left(1+\frac{m}{2r}\right)^4\left(dr^2+r^2(d\theta^2+\sin^2\theta\, d\phi^2)\right).
\end{equation}

\noindent This line element is a solution of the Einstein field equations \emph{only} for $r \neq m/2$, where $m/2$ is the location of the event horizon in isotropic coordinates, whereas this solution is valid \emph{everywhere} from the point of view of shape dynamics. The line element is invariant under the transformation  $r\to m^2/(4r)$. It can be seen from this inversion symmetry is that the ``interior" of this solution is a time-reversed copy of the exterior, which makes manifest the wormhole character of the solution and that the solution possesses no central curvature singularity. In fact, from the point of view of shape dynamics, this solution is completely free of physical singularities since there are no singularities in the spatial conformal geometry. This can be seen by noting that the Cotton tensor\footnote{The Cotton tensor is defined by $\mathscr{C}_{ijk} := \nabla_k\left(R_{ij} - \frac{1}{4}Rq_{ij}\right) - \nabla_j\left(R_{ik} - \frac{1}{4}Rq_{ik}\right)$. The Cotton tensor possesses all of the local information on the conformal structure of a three-dimensional Riemannian manifold just as the Weyl tensor contains all of the local information on the conformal structure of higher-dimensional manifolds \cite{Gourgoulhon}.} vanishes identically, which follows from the fact that the spatial metric is conformally flat.

\begin{figure}[h]
\begin{subfigure}{.5\textwidth}
  \centering
  \includegraphics[width=.8\linewidth]{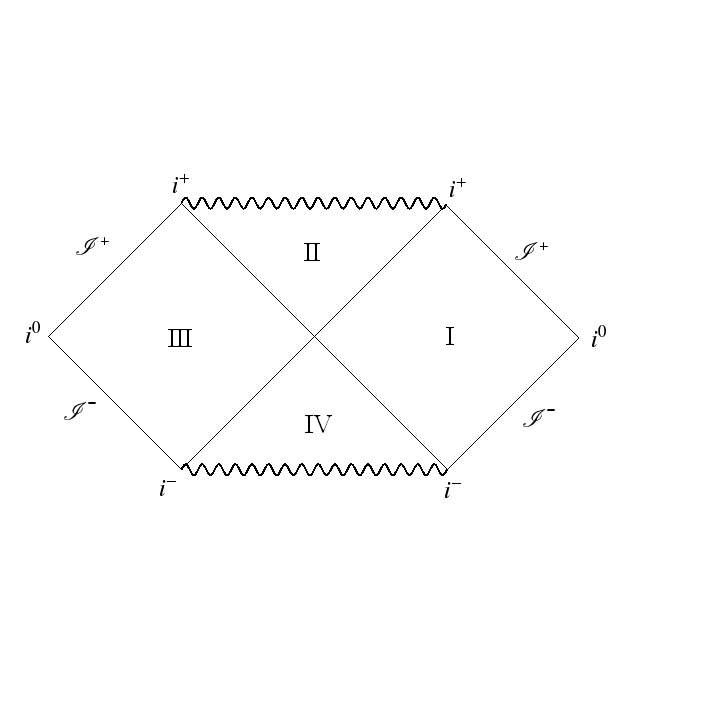}
  \caption{\,}
  \label{penrose-GR}
\end{subfigure}%
\begin{subfigure}{.5\textwidth}
  \centering
  \includegraphics[width=.8\linewidth]{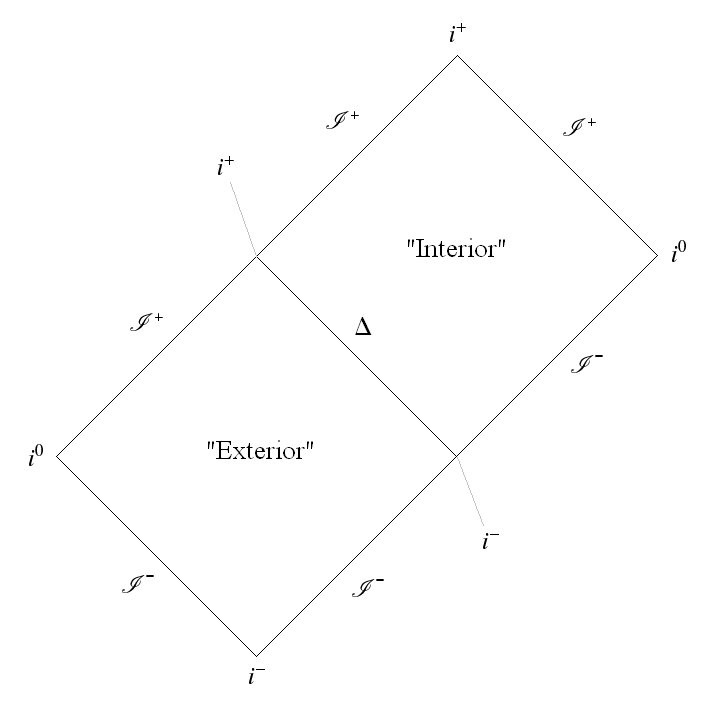}
  \caption{\,}
  \label{penrose-SD}
\end{subfigure}
\caption{On the left, (a) shows the conformal diagram of the maximally extended Kruskal spacetime. On the right, (b) shows the conformal diagram for a spherically symmetric, shape dynamic black hole. In each, $i^{\pm}$ are future and past null infinity, $i^0$ is spatial infinity, $\mathscr{I}^{\pm}$ are future and past null infinity and $\Delta$ is the event horizon.}
\label{penrose}
\end{figure}

This solution can be understood in relation to the maximally extended Kruskal spacetime by considering the conformal diagrams displayed in \ref{penrose}. The conformal diagram of the Kruskal spacetime (Fig. \ref{penrose-GR}) contains four regions. Regions I and III are causally disconnected and each possesses its own spatial infinity. Region II containing a future space-like singularity and region IV contains a past space-like singularity. On the other hand, the conformal diagram of the spherically symmetric shape dynamic black hole (Fig. \ref{penrose-SD}) contains only the two regions labeled (somewhat arbitrarily) ``interior" and ``exterior." Much like regions I and III of the conformal diagram of the Kruskal spacetime, the interior and exterior regions each possess their own spatial infinity. However, unlike the Kruskal spacetime, the interior and exterior regions meet along the event horizon $\Delta$, and are therefore not causally disconnected. Observers can pass in one direction across the horizon from the exterior region to the interior. The regions containing the past and future singularities are absent in the shape dynamics case, reflecting the global regularity of the solution.

The inversion symmetry of this black hole solution is known and has been discussed in \cite{Birkhoff} and \cite {Kerr}. This property can be used to motivate the following
\\
\begin{quote}
\noindent Definition: Let $\{N, \xi, q_{ij}, \pi^{ij}\}$ be a solution of the equations of motion of shape dynamics in the gauge in which $\rho(x) = 0$, $e^{4\phi(x)} = 1$ and let $\dot{q}_{ij} = 0$. Suppose $q_{ij}$ admits a two-surface $\mathcal{S}_0$ on which the lapse function $N(x)$ either vanishes or diverges and if there exists a spatial isometry $\mathcal{P}$ such that:
\begin{enumerate}
\item
 $\mathcal{P}^*q = q$ 
\item
 $\mathcal{P}\circ\mathcal{P} = \mathbb{1}$ 
\item
 $\mathcal{P}^*N = -N$.
\end{enumerate}
Then  $\mathcal{P}$ is called a parity and $\mathcal{S}_0$ is called a parity horizon. \\
\end{quote}

It is possible that this definition might be improved upon by restating it in a gauge invariant manner and/or extending it to include solutions with $\dot{q}_{ij} \neq 0$. However, for our purposes the above definition is adequate. Clearly, the event horizon of the spherically symmetric black hole solution discussed above is a parity horizon. In the remainder of this chapter, I will consider various examples of known and novel solutions of shape dynamics which possess parity horizons and discuss how the presence of this novel feature leads to physical differences between solutions of shape dynamics and solutions of general relativity that agree outside of horizons but disagree at and within them.

Let us now consider another shape dynamic black hole solution whose event horizon is a parity horizon. It was shown in \cite{Kerr} that shape dynamics admits a rotating black hole solution that also possesses an inversion symmetry about the horizon manifesting its wormhole character, and it was argued that this solution is also free of physical singularities. The reconstructed line element associated with this solution is reproduced here for the reader's convenience:
\begin{equation}\label{Kerr PS2} 
ds^2 = -\lambda^{-1}\left(dt - \omega^{-1}d\phi\right)^2 + \lambda \left[m^2e^{2\gamma}(d\mu^2 + d\theta^2) + s^2d\phi^2\right]
\end{equation}
where 
\begin{eqnarray}\label{Kerr functions PS} 
s \hspace{5pt} &=& mp\sinh\mu\sin\theta \nonumber \\
e^{2\gamma} &=& p^2\cosh^2\mu  + q^2\cos^2\theta - 1 \nonumber\\
\omega^{-1} \hspace{5pt} &=& e^{-2\gamma}\left[2mq\sin^2\theta(p\cosh\mu + 1)\right]  \\
\lambda \hspace{5pt} &=& e^{-2\gamma}\left[(p\cosh\mu + 1)^2 + q^2\cos^2\theta \right]. \nonumber \\ 
p^2 + q^2 &=& 1 \nonumber
\end{eqnarray} 
The solution becomes spherically symmetric when $p \to 1$, $q \to 0$. The location of the horizon in these coordinates is $\mu = 0$ and the transformation $\mu \to -\mu$ leaves the line element invariant and flips the sign of the lapse, so this solution also possesses a parity about the horizon. In addition to avoiding the ring curvature singularity that is present in the interior of the Kerr spacetime, this solution also avoids the closed timelike curves present in that region. Since shape dynamics also has a preferred time variable, it is natural to wonder if it is a general feature of solutions of shape dynamics that closed timelike curves do not occur. This may indeed be the case, and in chapter \ref{bonner} I will show that this is true for another solution of shape dynamics whose general relativistic counterpart contains closed timelike curves but possesses no event horizon. I will argue based on the generic features of this solution that chronology protection may be a general feature of the theory.

In the following section, I introduce a charged black hole solution for shape dynamics analogous to the Reissner-N\"{o}rdstrom black hole in general relativity. It is shown that the event horizon is again a parity horizon and that the parity about the horizon leads to a fully CPT invariant solution.

\section{A Charged Shape Dynamic Black Hole}\label{charge}
To address the question of whether charged shape dynamic black holes possess some notion of CPT invariance, one must first couple the linking theory to the electromagnetic field and perform a phase space reduction to obtain shape dynamics coupled to electromagnetism \cite{H-thesis}. The Hamiltonian density for electromagnetism is given by 
\begin{equation}\label{emHam}
\mathcal{H}_{\mbox{\tiny EM}} = 2\left(A_{[i,j]}A_{[k,l]}q^{ik}q^{jl}\sqrt{q} + \frac{\tilde{E}^i\tilde{E}^jq_{ij}}{\sqrt{q}}\right)
\end{equation}

\noindent where $A_i$ is the vector potential and $\tilde{E}^i$ is its canonically conjugate momentum. The physical electric field $E^i$ is related to the vector density $\tilde{E}^i$ by $E^i = \frac{\tilde{E}^i}{\sqrt{q}}$. Coupling electromagnetism to general relativity yields the system of constraints:
\begin{eqnarray}\label{coupled constraints}
\mathcal{S}(x) &=& \frac{\pi^{ij}\pi_{ij} - \pi^2}{\sqrt{q}} - \sqrt{q}R + \mathcal{H}_{\mbox{\tiny EM}} \\
\mathcal{H}_a(\xi^a) &=& \int d^3x\left(q_{ij}\mathcal{L}_{\xi}\pi^{ij} + A_{i}\mathcal{L}_{\xi}\tilde{E}^i\right) \\
G(x) &=& \nabla_i\tilde{E}^i 
\end{eqnarray}

The next step is to extend the phase space and perform the canonical transformation of the constraints. Before proceeding, one must determine how $A_i$ and $E^i$ transform under the canonical transformation $T_{\phi}$. Here it is assumed as in \cite{H-thesis}, that $A_i$ and $E^i$ transform trivially under $T_{\phi}$ so that one can retain a system of first class constraints with well-defined global charges.  With this additional assumption, the transformed coupled scalar constraint becomes
\begin{equation}\label{coupled LY-1}
T_{\phi}\mathcal{S}(x) = \sqrt{q}\Omega\left(8\nabla^2\Omega - R\right) + \frac{\pi_{ij}\pi^{ij}-\pi^2}{\sqrt{q}}\Omega^{-7} + \Omega^{-2}\mathcal{H}_{\mbox{\tiny EM}} \approx 0.
\end{equation}

\noindent where $\Omega=e^{\phi}$. Imposing the gauge-fixing condition $\pi_{\phi} \equiv 0$ one obtains the coupled second class constraint 

\begin{equation}\label{coupled LY}
T_{\phi}\mathcal{S}(x) \approx \sqrt{q}\Omega\left(8\nabla^2\Omega - R\right) + \frac{\pi_{ij}\pi^{ij}}{\sqrt{q}}\Omega^{-7} + \Omega^{-2}\mathcal{H}_{\mbox{\tiny EM}} \approx 0.
\end{equation}

\noindent where $\pi(x)$ is now once again viewed as a first class constraint generating spatial Weyl transformations. As in the uncoupled case, one can also derive a consistency condition that fixes the lapse function:
\begin{equation}\label{coupled LFE}
T_{\phi}\{\mathcal{S}(N), \pi(x)\} \approx
e^{-4\phi}(\nabla^2 N + 2q^{ij}\phi_{,i}N_{,j}) -
N\left(e^{-12\phi}G_{ijkl}\frac{\pi^{ij}\pi^{kl}}{|q|} + e^{-8\phi}\frac{\mathcal{H}_{\mbox{\tiny EM}}}{\sqrt{q}}\right) \approx 0.
\end{equation}

In order to find a simple solution to the coupled constraints, we consider spherically symmetric conformal initial data which trivially satisfy the coupled first class constraints if we choose $\rho = 0 = \xi^i$:
\begin{equation}\label{initial data}
\bar{q}_{ij} = \eta_{ij}, \hspace{.5cm} \bar{\pi}^{ij} = 0, \hspace{.5cm}\bar{A_i} = 0, \hspace{.5cm} \bar{E}^i = \frac{Q}{r^2}\hspace{.05cm}\delta^i_r
\end{equation}

\noindent where $\eta_{ij}$ is the flat spatial metric written in spherical coordinates, and the barred quantities are arbitrarily rescaled according to their conformal weight by a conformal factor $\Omega$ which is to be determined by solving \eqref{coupled LY}. Note that the chosen initial data is static and spherically symmetric, so the equations of motion are trivial, and that the spherically symmetric electric field $\bar{E}^i$ is just the Coulomb electric field in the flat background defined by $\bar{q}_{ij} = \eta_{ij}$.

Since the conformal initial data is written in terms of the flat spatial metric $\eta_{ij}$, the scalar curvature $R$ vanishes, and \eqref{coupled LY} becomes simply
\begin{equation}\label{LY temp1}
8\Omega^3\nabla^2\Omega + \frac{\mathcal{H}_{\mbox{\tiny EM}}}{\sqrt{q}} = 0.
\end{equation}

\noindent Putting \eqref{initial data} into \eqref{emHam} gives an explicit expression for $\mathcal{H}_{\mbox{\tiny EM}}$ which can be substituted into  \eqref{LY temp1} to give
\begin{equation}\label{LY temp2}
8\Omega^3\left(\Omega'' + \frac{2}{r}\Omega'\right) + \frac{2Q^2}{r^4} = 0
\end{equation}

\noindent where I have also used spherical symmetry to assume that $\Omega$ depends only on $r$ and primes denote differentiation with respect to $r$. Equation \eqref{LY temp2} is difficult to solve in its present form, but it can be simplified by making the substitution $\Omega^2 = \psi$, which yields
\begin{equation}\label{LY psi}
-2\left(\psi'\right)^2 + 4\psi\psi'' + \frac{8}{r}\psi\psi' + \frac{2Q^2}{r^4}=0.
\end{equation}

\noindent Now equation \eqref{LY psi} can be solved by making the Laurent series ansatz:

\begin{equation}\label{laurent}
\psi = \sum\limits_{n=0}^{\infty}c_nr^{-n}.
\end{equation}

\noindent The derivatives of $\psi$ can be easily calculated from \eqref{laurent}:
\begin{eqnarray}\label{laurent prime}
\psi' &=& -\sum\limits_{n=0}^{\infty}nc_nr^{-(n+1)} \nonumber \\
\psi'' &=& \sum\limits_{n=0}^{\infty}n(n+1)c_nr^{-(n+2)}.
\end{eqnarray}

\noindent Inserting \eqref{laurent} and \eqref{laurent prime} back into \eqref{LY psi} yields the infinite double sum:
\begin{equation}\label{double sum}
\sum\limits_{m=0}^{\infty}\sum\limits_{n=0}^{\infty}c_m c_n \left[-2mn + 4n(n-1)\right] r^{-(2 + m + n)} = -\frac{2Q}{r^4}.
\end{equation}

\noindent In order for \eqref{double sum} to be satisfied to all orders, all terms for which $m+n \neq 2$ on the left-hand side must vanish. This implies that $c_i = 0$ for $i>2$, which means that the series terminates. Collecting the terms proportional to $r^{-4}$ on the left hand side gives:
\begin{equation}\label{constants}
Q^2 + 4c_0 c_2 - c_1^2 = 0.
\end{equation}

\noindent Imposing the generic boundary conditions\footnote{The choice to rename $c_1 = m$ can be justified a fortiori by noting that the ADM mass of the resulting solution is equal to m.} $c_0 = 1$, $c_1 = m$, equation \eqref{constants} can be solved for $c_2$ in terms of the mass $m$ and electric charge $Q$:
\begin{equation}\label{c2}
c_2 = \frac{m^2-Q^2}{4}.
\end{equation}

\noindent Inserting equation  \eqref{c2} back into equation \eqref{laurent} and recalling that $c_n = 0$ for $n>2$:
\begin{equation}\label{Omega final}
\psi = 1 + \frac{m}{r} + \frac{m^2-Q^2}{4r^2} \hspace{.45cm}
\implies \hspace{.45cm}
\Omega =  \left(1 + \frac{m}{r} + \frac{m^2-Q^2}{4r^2}\right)^{1/2}.
\end{equation}

Putting equations \eqref{Omega final} and \eqref{initial data} into \eqref{coupled LFE}, one obtains the homogenous, linear, second order ordinary differential equation:
\begin{equation}\label{coupled LFE 2}
\Omega^4\left(N'' + 2\left(\frac{1}{r} + \frac{\Omega'}{\Omega\hspace{.07cm}}\right)N'\right) - \frac{Q^2}{r^4}N = 0
\end{equation}

\noindent which given asymptotically flat boundary conditions, has the unique solution:

\begin{equation}
N = \frac{1 - \frac{m^2 - Q^2}{4r^2}}{1 + \frac{m}{r} + \frac{m^2-Q^2}{4r^2}}.
\end{equation}

Rescaling the conformal initial data by the appropriate powers of $\Omega$ according to their conformal weights gives the spatial metric and physical electric field:
\begin{eqnarray}\label{rescaled initial data}
q_{ij} &=& \Omega^4\bar{q}_{ij} =  \left(1 + \frac{m}{r} + \frac{m^2-Q^2}{4r^2}\right)^2\eta_{ij} \\
E^i &=& \Omega^{-6}\bar{E}^i =  \left(1 + \frac{m}{r} + \frac{m^2-Q^2}{4r^2}\right)^{-3}\frac{Q}{r^2}\delta^i_r
\end{eqnarray}

The reconstructed line element associated with this solution of shape dynamics is given by
\begin{eqnarray}\label{Nord element}
ds^2 &=& -N^2 dt^2 + q_{ij}dx^i dx^j \nonumber \\ &=& \left(\frac{1 - \frac{m^2 - Q^2}{4r^2}}{1 + \frac{m}{r} + \frac{m^2-Q^2}{4r^2}}\right)^2 dt^2 + \left(1 + \frac{m}{r} + \frac{m^2-Q^2}{4r^2}\right)^{2}\left(dr^2 + r^2 dS_{\mbox{\tiny 2}}^2\right)
\end{eqnarray}

\noindent where $dS_{\mbox{\tiny 2}}^2 = d\theta^2 + \sin^2\theta d\phi^2$ is the metric on the unit two-sphere. Equation \eqref{Nord element} is just the line element of the Reissner-N\"{o}rdstrom black hole written in isotropic coordinates, which is well known in the context of general relativity, and has been derived by similar methods in for example \cite{Alcubierre}. The physical difference between this shape dynamics solution and the Reissner-N\"{o}rdstrom black hole is once again that the shape dynamics solution has a wormhole character manifested by the presence of a parity horizon. The event horizon of this black hole solution is located at $r = r_* = \frac{1}{2}\sqrt{m^2-Q^2}$ where the lapse vanishes. In the limit $Q \to 0$, one recovers the uncharged spherically symmetric shape dynamic black hole and the location of the horizon reduces to $r = r_* = m/2$. Just as in the uncharged spherically symmetric black hole, the spatial metric is invariant under the parity $r \to \tilde{r} = r_*^2/r$, while the lapse changes sign under this transformation. Thus, the surface $r = r_*$ is a parity horizon. 

It is interesting to consider how the electric field $\vec{E} = E^i\partial_i$ transforms under parity. One can check that 
\begin{equation}\label{E parity}
\vec{E} = \left(1 + \frac{m}{r} + \frac{m^2-Q^2}{4r^2}\right)^{-3}\frac{Q}{r^2}\partial_r \hspace{.3cm} \to  \hspace{.3cm}
 \left(1 + \frac{m}{\tilde{r}} + \frac{m^2-Q^2}{4\tilde{r}^2}\right)^{-3}\frac{Q}{\tilde{r}^2}\partial_{\tilde{r}} = -\vec{E}
\end{equation}

\noindent so the electric field transforms like a vector under parity and transforms trivially under time reversal $t \to -t$. The lapse changes sign under time-reversal, which can be seen by noting that N can be defined through the linking theory as $N = -t^{\mu}n_{\mu}$ where $n_{\mu}$ is the unit normal\footnote{In the present setting, where the lapse can naturally take on both positive and negative values, the normal vector is taken to be fixed under time reversal.} to the maximal space-like hypersurface of constant $t$. Finally, the electric field changes sign under charge conjugation $Q \to -Q$ while the spatial metric and lapse are invariant under this transformation. Putting all this together, it is clear that the charged spherically symmetric  shape dynamic black hole solution derived above is CPT invariant.

It is worth noting that while the uncharged spherically symmetric shape dynamic black hole is fully (C)PT invariant, the rotating solution is not. It is easy to see that while the parity preserves the spatial metric and changes the sign of the lapse, the shift vector transforms like a pseudo-vector under parity (i.e. it does \emph{not} transform) while it changes sign under time-reversal. Thus, the combination of parity and time-reversal has the effect of changing the sign of the angular momentum. 

However, one can still recover PT invariance of the rotating shape dynamic black hole as an asymptotic symmetry which is approximate in the bulk but becomes exact on the horizon. This can be done by noting that the spacetime coordinate transformation $\tilde{t} =  t - \omega_0^{-1}\phi$ where $\omega_0$ is the angular velocity of the horizon measured at spatial infinity, preserves the Weyl constraint and is thus a residual gauge symmetry inherited from the linking theory. The effect of this transformation is that the new time coordinate $\tilde{t}$ is defined so that its associated coordinate derivative $\partial_{\tilde{t}} = \partial_t + \omega_0\partial_{\phi}$ is the stationarity Killing vector defining the bifurcate Killing horizon of the reconstructed spacetime. In terms of this adapted time coordinate, \eqref{Kerr PS2} becomes:
\begin{equation}\label{new Kerr}
ds^2 = -\lambda^{-1}\left[d\tilde{t} + \left(\omega_0^{-1}- \omega^{-1}\right) d\phi\right]^2 + \lambda \left[m^2e^{2\gamma}(d\mu^2 + d\theta^2) + s^2d\phi^2\right].
\end{equation}

It is easy to see that in these coordinates the shift vector vanishes on the horizon, and the time reversal $\tilde{t} \to -\tilde{t}$ combined with the parity $\mu \to -\mu$ is an asymptotic symmetry of the horizon. In the static cases, there is no difference between t and $\tilde{t}$ since the angular velocity of the horizon is zero. This suggests that in defining time-reversal for stationary shape dynamic black holes, one should take $\tilde{t} \to -\tilde{t}$ rather than $t \to -t$. Based upon these considerations, one would expect that a Kerr-Newman-like shape dynamic black hole would possess an asymptotic CPT invariance that becomes exact on the horizon, but this is left for future work. 

While all of the black hole solutions I have considered are electrovac solutions, there is reason to think that parity horizons might emerge in the near horizon limit of black hole solutions of the shape dynamics equations of that contain matter in the exterior region as well. Medved, Martin and Visser have studied static and spherically symmetric but otherwise generic ``dirty" black holes \cite{MMV} in the canonical formalism\footnote{By using a time function whose associated coordinate derivative coincides with the static timelike Killing vector field, the momentum conjugate to the spatial metric naturally vanishes ensuring the Weyl constraint is trivially satisfied.} by expanding in powers of the radial proper distance from the horizon. Their results show that the lapse and spatial metric define a parity horizon at least up to cubic order for lapse and up to quadratic order for the spatial metric.  It is noteworthy that this is a local construction and does not make any assumptions, such as asymptotic flatness, about boundary conditions at infinity.  

Next, I will show that Rindler space can be viewed as a solution of shape dynamics and that the Rindler horizon is also a parity horizon. 
\chapter{Rindler Space as a Solution of Shape Dynamics}\label{rindler}
\section{The Rindler Chart}
The Rindler chart represents a congruence of uniformly accelerating observers over a portion of the Minkowski spacetime. It is of particular interest because despite the fact that it is related to ordinary Cartesian coordinates in flat spacetime by a spacetime coordinate transformation, the Rindler chart contains an observer-dependent horizon which has a non-zero surface gravity, and hence a non-zero temperature, much like the event horizon of a black hole.\footnote{This is a physical effect: If one considers a Klein-Gordon field propagating in Minkowski space, one can identify the vacuum state with respect to inertial observers, but this state is not a vacuum with respect to accelerated observers---rather accelerated observers see a thermal bath of excitations of the Klein Gordon field with temperature equal to $\kappa/2\pi$ where $\kappa$ is the surface gravity of the Rindler horizon \cite{Lambert}. This is entirely analogous to what happens when one considers a scalar field in a black hole background--- initial asymptotic vacuum states are not final asymptotic vacuum states, and the result is that the black hole creates particles in the form of Hawking radiation \cite{HawkingRadiation}. The fact that accelerated observers in flat spacetime observe a thermal bath is known as the ``Unruh effect."}

In order to construct the Rindler chart, one can begin with Cartesian coordinates over Minkowski spacetime. The line element is simply:
\begin{equation}\label{Minkowski}
ds^2 = -dT^2 + dX^2 + dY^2 + dZ^2.
\end{equation}

\noindent If one then introduces the coordinate transformation
\begin{eqnarray*}
t &=& \frac{1}{\kappa}\tanh^{-1}\left(\frac{T}{X}\right) \label{t-def} \\
x &=& \sqrt{X^2-T^2} \label{x-def} \\
y &=& Y \\
z &=& Z
\end{eqnarray*}
\noindent one obtains the Rindler chart with the line element
\begin{equation}\label{Rindler-first}
ds^2  = -\kappa^2x^2dt^2 + dx^2 + dy^2 + dz^2.
\end{equation}

The congruence of uniformly accelerating observers see a horizon located at $x = 0$, which can be seen by noting the time-time component of the metric goes to zero there. It is worth noting that the coordinate transformation \eqref{x-def} defines $x$ only for $x > 0$. For this reason, the chart is often called the ``right Rindler wedge," and the left wedge must be defined separately with $x$ defined to have the opposite sign. Since this solution was obtained by performing a coordinate transformation that mixes space and time, it is not obvious that it is physically equivalent to the Minkowski spacetime from the point of view of shape dynamics; spacetime diffeomorphisms are not a gauge symmetry of shape dynamics, only spatial diffeomorphisms and spatial Weyl transformations are. Note however, that this coordinate transformation preserves the Weyl constraint, so Rindler space should indeed be a solution of shape dynamics. Next, I will show how Rindler space can be derived as a solution of shape dynamics from first principles.

\section{Rindler Space in Shape Dynamics}
As mentioned earlier, shape dynamics trades the refoliation invariance generated by the Hamiltonian constraint of general relativity for spatial Weyl invariance generated by a new Weyl constraint. One consequence of this symmetry trading is that the lapse function is not a Lagrange multiplier in shape dynamics, but must instead satisfy  the lapse-fixing equation \eqref{LFE1}, which is reproduced below for convenience. 
\begin{equation*}
e^{-4\phi}(\nabla^2 N + 2q^{ab}\phi_{,a}N_{,b}) -
e^{-12\phi}G_{abcd}\frac{\pi^{ab}\pi^{cd}}{|q|}N = 0.
\end{equation*}
Now choose the flat initial data $q_{ij} =
\delta_{ij}$, $\pi^{ij} = 0$, and choose the gauge $\phi = 0$, so that the spatial slices in the reconstructed spacetime are flat, as opposed to conformally flat as they are in the spherically symmetric black hole cases. If the shift vector is chosen to be zero, then the reconstructed line element associated with this solution of shape dynamics will have the form:
\begin{equation}\label{static}
ds^2 = -N^2dt^2 + q_{ij}dx^idx^j.
\end{equation}

In this simple case, the lapse-fixing equation \eqref{LFE1} reduces to Laplace's equation, $\nabla^2N
= 0$. The general solution of the lapse-fixing equation for this initial data is now trivially given by the harmonic functions. In spherical coordinates, this yields:

\begin{equation}\label{harmonics}
N(r,\theta,\phi) = \sum\limits_{\ell=0}^{\infty}\sum\limits_{m=-\ell}^{\ell}\left(A_l\hspace{.05cm}r^{\ell} + B_{\ell}\hspace{.05cm}r^{-(\ell+1)} \right)Y_{\ell}^m(\theta,\phi)
\end{equation}

\noindent where $Y_{\ell}^m(\theta,\phi)$ are spherical harmonics. First, consider asymptotically flat boundary
conditions $N(\infty) = 1$ and $\partial_rN \sim \mathcal{O}(r^{-2})$, where the latter condition essentially imposes that the total energy of the solution is zero. Obviously, the only way that the first boundary condition can be satisfied is if $\ell=0$, so the solution becomes

\begin{equation}\label{harmonics plus BCs}
N(r,\theta,\phi) = \left(A_0 + \frac{B_0}{r} \right)Y_0^0(\theta,\phi).
\end{equation}

But $Y_0^0(\theta,\phi)$ is just a constant, and the second boundary condition implies that $B_0 = 0.$ Putting this together, we have $N=const.$, and since $N(\infty)=1$, this means $N=1$ everywhere. Switching back to Cartesian coordinates, one obtains the Minkowski line element:
\begin{equation}\label{minkowski}
ds^2 = -N^2dt^2 + q_{ij}dx^idx^j = -dt^2 + dx^2 + dy^2 + dz^2. \\
\end{equation} 

On the other hand, one can consider Cartesian spatial coordinates, and demand that $N(x=0) = 0$, $\frac{dN}{dx}\big|_{x=0}
= \kappa$, where $\kappa$ is a constant.\footnote{It is not necessarily obvious that such boundary conditions should be permissible. Usually, in an asymptotically flat setting, one would place boundary conditions on the lapse at infinity and it is not immediately obvious that the ``horizon" boundary conditions we have chosen, will translate into a physically acceptable solution. However, this choice is justified a fortiori, by noting that the resulting solution for the lapse grows as the first power of $r$, leading to a finite (zero) energy for the solution.} It is interesting to note that this problem can be mapped onto the
problem of finding the electric potential (for $x>0$) due to an
infinite plane of surface charge at $x = 0$, so one can more or less guess the solution $N = Ex$, where is $E$ is some constant playing the role of the electric field. It is a simple exercise to show that the constant playing the role of the electric field is exactly $\kappa$, and this is carried out in appendix C for the sake of completeness.

The reconstructed line element has the form of the
Minkowski spacetime written in Rindler coordinates:

\begin{equation}\label{Rindler}
ds^2 = -N^2dt^2 + q_{ij}dx^idx^j = -\kappa^2x^2dt^2 + dx^2 + dy^2 + dz^2.
\end{equation}

It is not terribly surprising that non-inertial
observers can be obtained by changing the boundary conditions on the lapse; Eulerian
observers\footnote{Eulerian observers are congruences of timelike curves whose tangent vectors are orthogonal to the spatial hypersurfaces of a (reconstructed) foliation of spacetime. These are natural observers to choose from the point of view of shape dynamics as they are ``inertial" with respect to the spatial geometry (hypersurface orthogonal), although they are generally non-inertial in that their proper acceleration is non-zero in the reconstructed spacetime.} are generally non-inertial, with their proper acceleration
given by $a_i =\nabla_i (\ln N)$. What is more surprising is that the
spacetime reconstructions obtained from these different boundary
conditions on the lapse are related by a spacetime
diffeomorphism. The fact that this is a \emph{residual} gauge transformation from the point of view of shape dynamics---i.e. a large gauge transformation inherited from the linking theory---may shed some light on the fact that the accelerated Eulerian observers associated with the Rindler chart observe physically different effects than their non-accelerated counterparts in the usual Cartesian Minkowski chart--- they observe a thermal bath of Unruh radiation.  Residual spacetime coordinate invariance in shape dynamics was partially explored in \cite{Lorentz}, but it is still not completely understood what role these transformations play in the theory. From the point of view of shape dynamics, these symmetries, which as in the present case may be disconnected from the identity, seem somewhat less fundamental than the gauge transformations generated  by the first class constraints which make no reference to any properties of particular solutions, and are pure-gauge even off-shell. 

Finally, it is worth noting that while at a glance the line elements seem to agree, the Rindler chart is a complete solution from the point of view of shape dynamics. That is, the coordinate $x$ is well defined for all real values and therefore covers both the right and left wedges. As a solution of shape dynamics, the Rindler chart is therefore globally different from the corresponding solution of general relativity where it is defined only for the right and left wedges separately, and globally also includes future and past wedges that are not present in the shape dynamics solution. These global differences are entirely analogous to those found when comparing the black hole solutions of shape dynamics and general relativity, but in the case of the Rindler chart, there is no event horizon. It can be easily seen that the spatial metric is invariant under the parity $x \to -x$, while the lapse changes sign under this parity, just as in the black hole solutions. Thus, the observer-dependent Rindler horizon is a parity horizon from the point of view of shape dynamics. This suggests that global disagreement with shape dynamics may arise whenever a stationary horizon of \emph{any} kind is present in a solution general relativity, and that the corresponding solution of shape dynamics will contain a parity horizon.

In the following section, I will present yet another solution of shape dynamics that possesses a parity horizon and that presents physical differences from the corresponding solution of general relativity in that the general relativistic solution contains closed timelike curves while the shape dynamics solution does not. This can be seen as an advantage for shape dynamics, as closed timelike curves violate causality and solutions of general relativity containing closed timelike curves are generally explained away on a case by case basis as being ``unphysical" for various reasons (e.g. they are quantum-mechanically unstable or they require non-localized matter distributions). In shape dynamics, no such ad hoc apologies are necessary---solutions of shape dynamics simply seem not to contain these pathologies.
\chapter{The Bonner Spacetime as a Solution of Shape Dynamics}\label{bonner}

\section{The Van Stockum Bonnor spacetimes}

The Van Stockum-Bonnor spacetimes are a class of asymptotically flat stationary, axisymmetric solutions of Einstein's equations sourced by rigidly rotating dust. The solutions are ultra-relativistic in the sense that they possess no Newtonian limit.\footnote{(There is an exception to this claim. The lowest order case is invariant under translations parallel to the $z$-axis, and is therefore \emph{not} asymptotically flat.} Indeed, it was pointed out in \cite{Bonnor} that any density gradient in the $z$-direction would produce gravitational forces parallel to the $z$-axis that would violate the assumption of rigid rotation. In general relativity, solutions with rigidly rotating dust are possible provided the angular velocity of the dust is sufficiently rapid.

Let us briefly review the properties of the Van Stockum-Bonnor spacetimes. Our discussion closely follows that of \cite{VanStockum} which the interested reader may consult for further details.  In cylindrical coordinates, the Bonnor-Van Stockum spacetimes are described by the line element
\begin{equation}\label{Van Stockum}
ds^2 = -dt^2 + 2K(\rho,z)d\phi\,dt + \left(\rho^2-K^2(\rho,z)\right) d\phi^2 + e^{2\Psi(\rho,z)}\left(d\rho^2 + dz^2 \right).
\end{equation} 

\noindent The Einstein field equations imply that
\begin{eqnarray}
\Psi_{,\rho} &=& \frac{K_{,z}^2 - K_{,\rho}^2}{4\rho} \label{EFE1} \\
\Psi_{,z} &=& -\frac{K_{,\rho}K_{,z}}{2\rho} \label{EFE2}
\end{eqnarray} 

\noindent Differentiating \eqref{EFE1} and \eqref{EFE2} with respect to $z$ and $\rho$ respectively, and noting that the partial derivatives commute then requires that $K(\rho,z)$ satisfy the linear, elliptic partial differential equation:
\begin{equation}\label{K-PDE}
K_{,\rho\rho} - \frac{1}{\rho}K_{,\rho} + K_{,zz} = 0.
\end{equation}

Given a solution of \eqref{K-PDE}, one can solve the system \eqref{EFE1}, \eqref{EFE2}  for $\Psi$ which completely determines the line element \eqref{Van Stockum}. On the other hand, it was shown in \cite{Kerr} that the general line element (up to coordinate transformations) representing a stationary, axisymmetric solution of the Einstein field equations 
\begin{equation}\label{Axi}
ds^2 = -(N^2 - \Omega\Phi\xi^2)dt^2 + \Omega[(dx^1)^2 + (dx^2)^2 + \Phi d \phi^2] + 2\Omega\Phi\xi d \phi\,dt
\end{equation}

\noindent induces a maximal slicing by hypersurfaces of constant $t$. Putting $x^1 = \rho$, $x^2 = z$, \eqref{Axi} can be put in the form of \eqref{Van Stockum} by making an appropriate choice of the metric functions $N$, $\Omega$, $\Phi$, and $\xi$:
\begin{equation}\label{metric functions}
N = \left(1 - K^2/\rho^2\right)^{-1/2}, \hspace{.5cm}
\Omega = e^{2\Psi}, \hspace{.5cm}
\Phi = e^{-2\Psi}\left(\rho^2 - K^2\right), \hspace{.5cm}
\xi = \frac{K}{\rho^2-K^2}.
\end{equation}

 This implies that \eqref{Van Stockum} induces a maximal slicing by hypersurfaces of constant $t$, which is precisely the criterion needed to map a solution of general relativity onto a solution of shape dynamics, at least locally.  Every solution of \eqref{K-PDE} yields a solution of \eqref{Van Stockum} that possesses a surface defined by the equation $\rho^2 = K^2$, and it can be readily seen that such a surface has two important properties. 

First, on such a surface $g_{\phi\phi}=0$, and it can be easily checked for any solution that this divides the spacetime into two regions: an ``exterior" region defined by $\rho^2>K^2$ and an ``interior region" defined by $\rho^2<K^2$. In the interior region, $g_{\phi\phi}<0$, and the coordinate basis vector $\partial_{\phi}$ is therefore \emph{timelike} in this region. Since the coordinate $\phi$ is periodic with period $2\pi$, the integral curves of $\partial_{\phi}$ are closed. Therefore, the interior region is filled with closed timelike curves. Any surface which divides a spacetime into a causal region without closed timelike curves, and an acausal region with closed timelike curves is a special kind of Cauchy horizon called a chronological horizon \cite{Thorne}.

The second important feature of this chronological horizon is that the lapse function, $N$ goes to infinity there, as can be easily from the first of equations \eqref{metric functions}. Recall that the determinant of the spacetime metric can be written $\sqrt{-g} = N\sqrt{q}$. If this quantity is finite, as it must be for a non-singular region of spacetime, then if the lapse function $N$ diverges on some surface, the determinant of the spatial metric must vanish. This is indeed the case, since $g_{\phi\phi}=q_{\phi\phi}=0$ and there are no off-diagonal terms in the spatial metric. This is a signal that the ``spatial" metric changes signature across this surface and is no longer truly spatial in the interior region. This is physically unacceptable from the point of view of shape dynamics, since shape dynamics is a theory of the time evolution of conformal equivalence classes of \emph{Riemannian} three-manifolds. Nevertheless, the fact that the line element \eqref{Van Stockum} induces a maximal slicing by hypersurfaces of constant $t$ in the exterior  region, means that this is a solution of shape dynamics in the exterior region. In what follows, I will consider a spatial coordinate transformation that is a diffeomorphism only in the exterior region, but which takes the line element \eqref{Van Stockum} into a form that makes the spatial metric non-degenerate across the chronological horizon, yielding a complete solution of shape dynamics that covers the whole spatial manifold. 

\section{The Bonnor Solution for Shape Dynamics}

Equation \eqref{K-PDE} is linear and invariant under translations in $z$-direction. Consequently, any linear combination of known solutions and their $z$-derivatives of any order is again a solution. A simple class of solutions can be obtained by expanding $K(\rho,z)$ in a superposition of modes. The external multipolar solutions are given by 

\begin{equation}\label{modes}
  K(\rho,z)=\begin{cases}
   z\left(\rho^2 + z^2\right)^{-\frac{n+1}{2}}\hspace{.22mm}_2F_1\left(\frac{n+1}{2},-\frac{n}{2};\frac{3}{2};\frac{z^2}{\rho^2+z^2}\right), & \text{$n=0,2,4,6...$}\\
\hspace{2.4mm}  \left(\rho^2 + z^2\right)^{-n/2}\hspace{1mm}_2F_1\left(-\frac{n+1}{2},\frac{n}{2};\frac{1}{2};\frac{z^2}{\rho^2+z^2}\right)  , & \text{$n=1,3,5,7...$}
  \end{cases}
\end{equation}

\noindent where $\hspace{.22mm}_2F_1(a,b;c;\eta)$ is the \emph{hypergeometric function} defined for $|\eta|<1$ by the infinite series

\begin{equation}\label{hypergeo}
\hspace{.22mm}_2F_1(a,b;c;\eta) = \sum\limits_{m=0}^{\infty}\frac{(a)_m(b)_m}{(c)_m}\frac{\eta^m}{m!}
\end{equation}

 \noindent and $(q)_m$ is the Pochhammer symbol:

\begin{equation}\label{weird thing}
(q)_m = \begin{cases}
1, &\text{$m=0$} \\
q(q+1)...(q+m-1), &\text{$m>0$}.
\end{cases}
\end{equation}

\noindent When either $a$ or $b$ is a non-positive integer, as it is in the expressions for the multipolar solutions for $K(\rho,z)$, the hypergeometric function becomes a polynomial of finite order:

\begin{equation}\label{hypergeo-finite}
\hspace{.22mm}_2F_1(-n,b;c;\eta) = \sum\limits_{m=0}^{n}(-1)^m\colvec{n}{m}\frac{(b)_m}{(c)_m}\eta^m.
\end{equation}  

\noindent For simplicity, I will consider a particular solution to \eqref{K-PDE} and its associated line element \eqref{Van Stockum} corresponding to the $n=1$ mode of \eqref{modes}, for which 

\begin{equation}
K(\rho,z) = 2h\frac{\rho^2}{\left(\rho^2 + z^2\right)^{3/2}}
\end{equation}

 and 

\begin{equation}
\Psi(\rho,z) = \frac{h^2}{4}\frac{\rho^2\left(\rho^2-8z^2\right)}{\left(\rho^2 + z^2\right)^4}
\end{equation}

\noindent where $h$ is a positive constant with dimensions of area parametrizing the location of the chronological horizon. Details about this solution can be found in \cite{Bonnor}. 

It is more convenient to work in spherical coordinates, where $K(r,\theta) = \frac{2h}{r}\sin^2\theta$, and \\ $\Psi(r,\theta) = \frac{h^2}{4}r^{-4}\sin^2\theta\left(\sin^2\theta - 8\cos^2\theta\right)$. After changing to spherical coordinates, the line element \eqref{Van Stockum} becomes
\begin{equation}\label{VanStockSphere}
ds^2 = -dt^2 +2Kd\phi dt + e^{2\Psi}\left(dr^2 + r^2d\theta^2\right) +\left(r^2\sin^2\theta-K^2\right)d\phi^2.
\end{equation}

This solution is known as the Bonnor spacetime, and in spherical coordinates, the equation defining the chronological horizon is given by

\begin{equation}\label{horizon}
 r^2\sin^2\theta - \frac{4h^2}{r^2}\sin^4\theta = g_{\phi\phi} = 0 \hspace{5mm}
\text{or} \hspace{5mm}
r^2 = 2h\sin\theta.
\end{equation}

\begin{figure}[h]
\centering
\includegraphics[width = 7.5cm, height = 5.5cm]{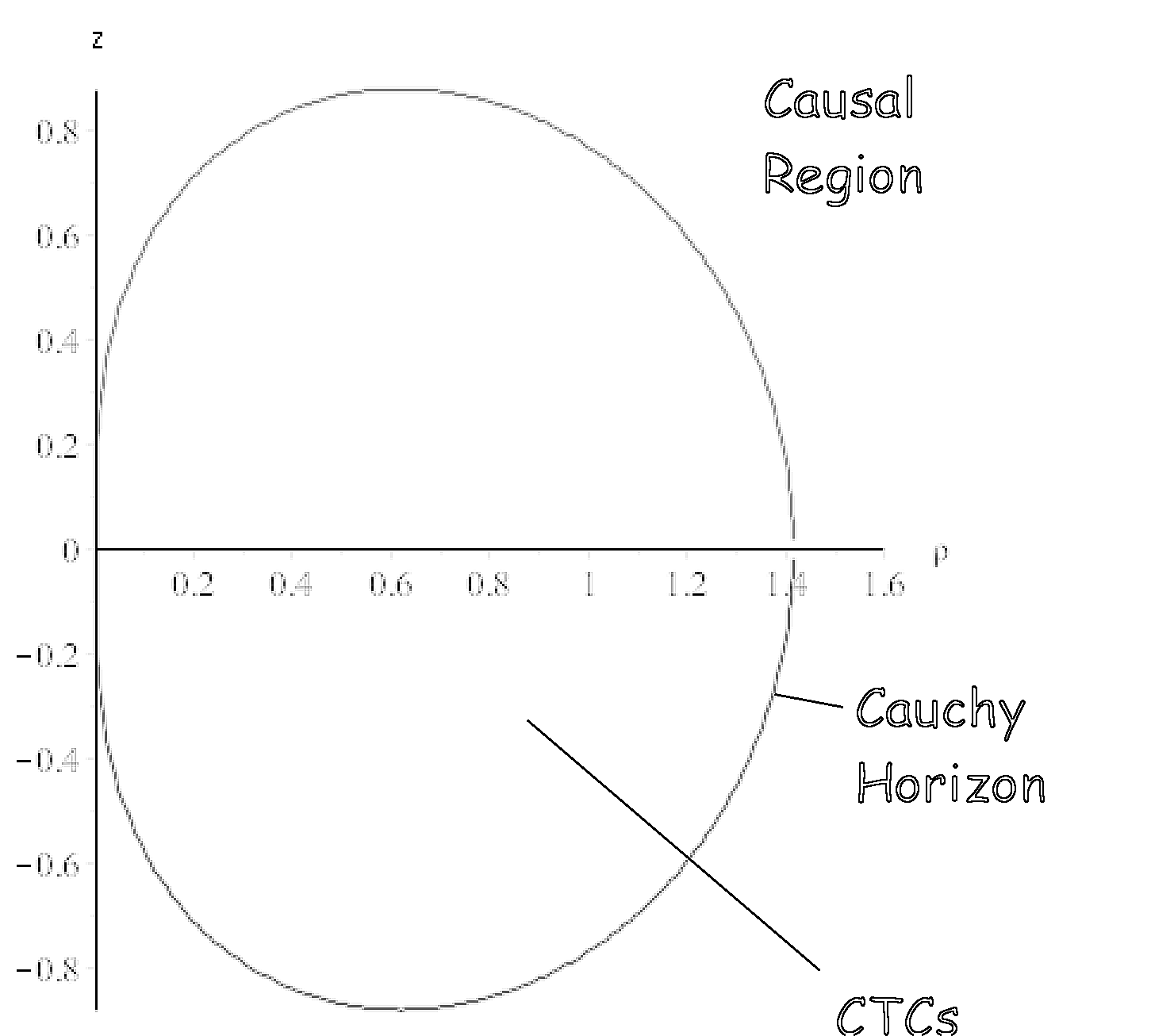} \\
\caption{A plot of the chronological horizon horizon in the Bonnor spacetime for $h = 1$.}
\label{rosy plane}
\end{figure}

The chronological horizon in the Bonnor spacetime has roughly the shape of a degenerate torus, with its inner circumference shrunken to a point. Figure \ref{rosy plane} depicts the chronological horizon in the $\rho\hspace{.3mm}$-$z$ half plane.

Equation \eqref{horizon} depends on both $r$ and $\theta$. To find coordinates that make the spatial metric smooth across the chronological horizon, it will be convenient to work in coordinates that are adapted to this surface, so that \eqref{horizon} takes the form $x = constant$ for some new spatial coordinate $x$. This can be accomplished by making the coordinate transformation
\begin{eqnarray}\label{to u and v}
x &=& \sqrt{r^2-2h\sin\theta} \\
y &=& \sqrt{r^2+2h\sin\theta}
\end{eqnarray}

\noindent In the $(x,y)$ system, the chronological horizon is located at $x=0$, and the spatial part of the line element becomes
\begin{eqnarray}\label{xy-element}
d\ell^2 &=& q_{ij}dx^idx^j \nonumber \\  &=& e^{2\Psi}\left[\frac{\frac{1}{2}\left(xdx+ydy\right)^2}{x^2+y^2} + \frac{1}{8h^2}\left(x^2+y^2\right)\frac{\left(y^2-x^2\right)^2}{\left(y^2-x^2\right)^2-16h^2}\left(ydy-xdx\right)^2\right] \nonumber \\ &\,& +\frac{x^2y^2}{32h^2}\frac{\left(y^2-x^2\right)^2}{x^+y^2}d\phi^2. 
\end{eqnarray}

\noindent In the near-horizon limit $x^2<<y^2$, \eqref{xy-element} reduces to

\begin{equation}\label{near-horizon}
d\ell^2 \approx e^{2\Psi}\left[Q_+(y)\left(x^2dx^2 + y^2dy^2\right) + 2Q_-(y)xy\,dx\,dy\right] + \frac{x^2y^4}{32h^2}d\phi^2
\end{equation}

\noindent where I have defined 

\begin{equation}\label{Q-def}
Q_{\pm}(y) = \frac{1}{2y^2} \pm \frac{1}{8h^2}\frac{y^4}{y^4-16h^2}.
\end{equation}

\noindent From equation \eqref{near-horizon} one can read off the determinant of the spatial metric in the near horizon limit:

\begin{equation}
|q| = e^{4\Psi}\left(Q_+^2-Q_-^2\right)\frac{x^4y^6}{32h^2}.
\end{equation}

Clearly, as $x \to 0$, $|q| \to 0$ as well, indicating that the spatial metric is still degenerate across the chronological horizon in the $(x,y)$ system. This can be remedied by making one final spatial coordinate transformation that is a diffeomorphism only for $x>0$. Since the determinant of the metric transforms as $|q| \to |\overbar{q}|=|J|^2|q|$ under a coordinate transformation $dx^b \to d\overbar{x}^a = J^a_b dx^b$ with Jacobian $J^a_b = \frac{\partial \overbar{x}^a}{\partial x^b}$, one can make the near-horizon spatial geometry non-degenerate by transforming the $x$ coordinate so that the $x$-dependence of the near-horizon determinant of the spatial metric is canceled. This can be done by choosing a new coordinate $u$ so that $\frac{du}{dx} \propto x^2$. Making an analogous transformation of the $y$ coordinate to preserve the symmetry of the metric, one can choose:
\begin{eqnarray}
u &=& h^{-1}x^3 \label{x to u}, \\
v &=& h^{-1}y^3 \label{y to v}.
\end{eqnarray}

\noindent where the prefactors of $h^{-1}$ appear so that the new coordinates have dimensions of length. In the $(u,v)$ system, the spatial line element becomes
\begin{eqnarray}\label{uv-element}
d\ell^2 &=& \overbar{q}_{ij}d\overbar{x}^id\overbar{x}^j \nonumber \\
&=& e^{2\Psi}\left[\overbar{Q}_+(u,v)\left(u^{-2/3}du^2 + v^{-2/3}dv^2\right) + 2\overbar{Q}_-(u,v)u^{-1/3}v^{-1/3}du\,dv\right] \nonumber \\  
&\,& + \frac{u^{2/3}v^{2/3}}{32}\frac{\left(v^{2/3}-u^{2/3}\right)^2}{u^{2/3}+v^{2/3}}d\phi^2
\end{eqnarray}

\noindent where I have defined 

\begin{equation}\label{Q-bar}
\overbar{Q}_{\pm}(u,v) = \frac{1}{18}\left[\frac{\alpha^2}{u^{2/3}+v^{2/3}} \pm \left(u^{2/3}+v^{2/3}\right)\frac{\left(v^{2/3}-u^{2/3}\right)^2}{\left(v^{2/3}-u^{2/3}\right)^2-16\alpha^2}\right], \hspace{3mm} \alpha = h^{1/3}. 
\end{equation}

\noindent In the near horizon limit $u^{2/3}<<v^{2/3}$, one finds that 

\begin{equation}\label{Q-bar limit}
\overbar{Q}_{\pm}(u,v) \approx \overbar{Q}_{\pm}(v) = \frac{1}{18}\left[\frac{\alpha^2}{v^{2/3}} \pm v^{2/3}\frac{v^{4/3}}{v^{4/3}-16\alpha^2}\right]
\end{equation}

\noindent and the near-horizon limit of the spatial determinant becomes 

\begin{equation}\label{near-horizon uv det} 
|\overbar{q}| \hspace{1mm} \approx \frac{e^{4\Psi}}{32}\left(\overbar{Q}_+(v)^2-\overbar{Q}_-(v)^2\right)\frac{\left(v^{2/3}-u^{2/3}\right)^2}{u^{2/3}+v^{2/3}}
\end{equation}

\noindent which is finite and non-zero on the chronological horizon $u=0$. 

Clearly, the transformation \eqref{x to u} is a diffeomorphism only for $u>0$, i.e. outside the chronological horizon, as its inverse transformation $x = (hu)^{1/3}$ is not differentiable at $u = 0$. Nevertheless, the spatial metric is non-degenerate across the chronological horizon, and the coordinate $u$ can be extended to all real values. Noting that the line element \eqref{uv-element} is invariant under $u \to -u$, it can easily be seen that the spatial geometry possesses a reflection symmetry about the chronological horizon. Moreover, in the $(u,v)$ system, the lapse becomes
\begin{equation}
N(u,v) = \frac{u^{2/3} + v^{2/3}}{(uv)^{1/3}}
\end{equation}

\noindent which changes sign under the parity $u \to -u$ . Just as in the other solutions of shape dynamics with horizons that have been considered, the ``interior" region may be interpreted as a time-reversed copy of the exterior region---the chronological horizon becomes a parity horizon. This makes it obvious that no portion of the reconstructed spacetime possesses closed timelike curves. Finally, it is worth emphasizing that since the spatial geometry is non-degenerate for all values of $u$, and since the line element \eqref{uv-element} is globally related to \eqref{Axi} by a spatial diffeomorphism (except on the horizon, which needs to be treated with some care), this line element corresponds to a \emph{complete} solution of shape dynamics, covering the entire spatial manifold except the horizon, $u = 0$ which will be discussed in the following section.

\section{Conformal Singularity of the Chronological Horizon}

In the previous section, it was shown that in the $(u,v)$ system the spatial metric \eqref{uv-element} becomes non-degenerate and the chronological horizon becomes a parity horizon. The canonical momentum in these coordinates can be found by considering Hamilton's equation for $\dot{q}_{ij}$\footnote{The bar over the metric has been omitted for notational convenience.}:
\begin{equation}\label{H1}
\dot{q}_{ij} = 4\rho(x) q_{ij} + 2e^{-6\phi(x)}\frac{N}{\sqrt{q}}\left(\pi_{ij} - \frac{1}{2}\pi q_{ij}\right) + \mathcal{L}_{\xi}q_{ij}.
\end{equation}

\noindent where $\mathcal{L}_{\xi}q_{ij}$ denotes the Lie derivative of the spatial metric along the shift vector. With the gauge fixing conditions $\rho(x) = 0$, $\phi(x) = 0$, and noting that $\dot{q}_{ij} = \partial_{\phi}q_{ij} = 0$, \eqref{H1} yields the simple expression
\begin{equation}\label{pi}
\pi_{ij} = -\frac{\sqrt{q}\hspace{.3mm}q_{\phi\phi}}{N}\left(\delta^u_{(i}\delta^{\phi}_{j)}\xi_{,u} + \delta^v_{(i}\delta^{\phi}_{j)}\xi_{,v}  \right)
\end{equation}

\noindent for the canonical momentum $\pi_{ij}$. It is easily seen from \eqref{H1} that the trace of the momentum $\pi = 0$ and this can be explicitly confirmed from \eqref{pi} and \eqref{uv-element}, so the Weyl constraint is satisfied. One can also check that as $u \to 0$, $\pi_{u\phi} \sim u^{-2/3} \to \infty$. This is not surprising as the transformation to the $(u,v)$ system deliberately introduced a singularity in the $q_{uu}$ and $q_{uv}$ components of the spatial metric in order to cancel the degeneracy from the vanishing the of $q_{\phi\phi}$ component. Next, it will be shown that the singularities appearing in spatial metric and its conjugate momentum are of a physical nature---i.e. the chronological horizon of the Bonnor solution for shape dynamics is an extended physical singularity. Finally, we argue that this result can generically be extended to a broader class of solutions containing chronological parity horizons. 

Since shape dynamics is a theory of evolving conformal geometries, our analysis will focus on invariants constructed from the conformally invariant degrees of freedom of the spatial curvature. In three spatial dimensions, the rank-three Cotton tensor

\begin{equation}\label{Cotton1}
\mathscr{C}_{ijk} := \nabla_k\left(R_{ij} - \frac{R}{4}q_{ij}\right) - \nabla_j\left(R_{ik} - \frac{R}{4}q_{ik}\right)
\end{equation}

 \noindent contains all of the local information concerning the conformal geometry of space with the Cotton tensor vanishing if and only if the spatial geometry is conformally flat. The algebraic complexity of the line element \eqref{uv-element} makes computation of the Cotton tensor difficult, but one can analyze the behavior of the spatial conformal geometry of the exterior by considering the Cotton tensor in spherical coordinates for $h=1$. When this is done, one can compute the square of the Cotton tensor:
\begin{equation}\label{C-squared spherical}
\mathscr{C}^{ijk}\mathscr{C}_{ijk} = 4r^{-22}\exp\left(\frac{3}{2}r^{-4}\sin^2\theta\left(3\cos\theta+1\right) \left(3\cos\theta-1\right)\right)\frac{A(r,\cos\theta)}{B(r,\cos\theta)}
\end{equation}

\noindent where $A(r,\cos\theta)$ is a polynomial of degree 32 in $r$ and $\cos\theta$ with integer coefficients and $B(r,\cos\theta)$ is a polynomial of degree 24 in $r$ and $\cos\theta$ with integer coefficients. It can be shown using numerical methods that $A(r,\cos\theta)$ is non-zero on the horizon, while $B(r,\cos\theta)$ goes to zero there. Plots of $B(r,\theta)$ are shown below. 

\begin{figure}[h]
\centering
\includegraphics[width = 7.5cm, height = 5.5cm]{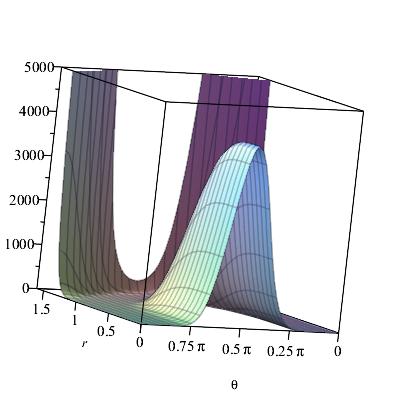} \\
\caption{A plot of $B(r,\theta) < 5000$ for $h = 1$ with $0 < r < 1.6$, $0 < \theta < \pi$.}
\label{3d1}
\end{figure}

\begin{figure}[h]
\centering
\includegraphics[width = 7.5cm, height = 5.5cm]{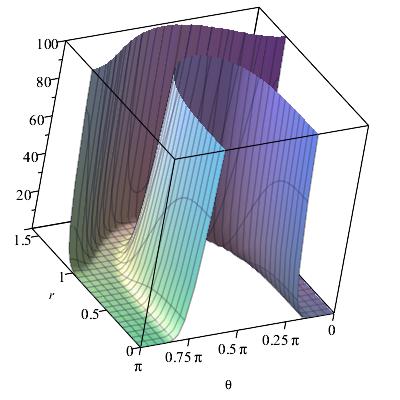} \\
\caption{A plot of $B(r,\theta) < 100$ for $h = 1$ with $0 < r < 1.6$, $0 < \theta < \pi.$}
\label{3d2}
\end{figure}

\ref{3d1} shows the general behavior of $B(r,\theta)$ near the horizon, while the range in \ref{3d2} is restricted to $[0,100]$ in order to display the shape of the horizon in the $r-\theta$ plane. One can see from figures 2 and 3 that $B(r,\theta)$ goes smoothly to zero along a curve in the $r-\theta$ plane, and it can be shown that this curve exactly coincides with the chronological horizon $r^2 = 2h\sin\theta.$ For values of $r$ and $\theta$ that lie in the interior of the horizon, the plots have no physical significance, as the $(r,\theta)$ system yields an ill-defined spatial conformally geometry for these values. Nevertheless, we can gain insight from the behavior of the invariant $\mathscr{C}^{ijk}\mathscr{C}_{ijk}$ outside and on the horizon. Since this invariant diverges smoothly as one approaches the horizon, and since it is an invariant of the spatial conformal geometry, there exists no spatial diffeomorphism of the exterior that can remove this divergence. Thus, we are forced to conclude that the cost of removing the degeneracy of the spatial metric in order to find a complete solution of shape dynamics is that the parity horizon of the shape dynamics solution must be regarded as an extended physical singularity. It is noteworthy that this is the first known solution of shape dynamics that possesses a ``shape singularity"---i.e. a singularity in the spatial conformal geometry.  

There is good reason to think that the feature we have just observed, namely that the chronological horizon becomes an extended physical singularity in shape dynamics, holds quite a bit more generally than in the specific case we have considered above. If we consider the spatial metric associated with the line element \eqref{Axi}, we can compute the general form of the invariant $\mathscr{C}^{ijk}\mathscr{C}_{ijk}$ in cylindrical coordinates:
\begin{eqnarray}\label{C-squared}
&\,&\mathscr{C}^{ijk}\mathscr{C}_{ijk} =  \nonumber \\ &\,& \frac{1}{ 4\psi^
{6} \Omega^{3}} 
\Bigg[ \psi^{2}{
\frac {\partial ^{3}\psi}{\partial {z}^{3}}} +
\psi^{2}{\frac {\partial ^{3}\psi}{
\partial {\rho}^{2}\partial z}}-2\,  \psi{\frac {\partial\psi }{\partial z}}{
\frac {\partial ^{2}\psi}{\partial {z}^{2}}}
 -\hspace{.5mm}2\, \psi {
\frac {\partial \psi}{\partial \rho}}{\frac {\partial ^{2}\psi}{\partial \rho
\partial z}}  + {\frac 
{\partial\psi }{\partial z}}\left( {
\frac {\partial\psi }{\partial \rho}}\right) ^{2}+
 \left( {\frac {\partial\psi }{\partial z}}
 \right) ^{3} \Bigg] ^{2} \nonumber \\ &+& (\rho \leftrightarrow z) 
\end{eqnarray}

Since the chronological horizon for any such solution is defined by $\psi(\rho,z)=0$, we see that the invariant generically diverges for any stationary, axisymmetric solution of shape dynamics with a chronological horizon so long as the numerator of the invariant does not conspire to cancel this divergence.

\chapter{Concluding Remarks}

The various known and novel stationary, axisymmetric solutions of asymptotically flat shape dynamics presented in this work demonstrate the pervasiveness of parity horizons in this context. The fact that black hole solutions of shape dynamics are physically different from those of general relativity opens up new possibilities for black hole physics and thermodynamics if shape dynamics is the correct description of the true degrees of freedom of the classical gravitational field. Some very preliminary results on the thermodynamics of shape dynamic black holes was discussed in \cite{Vasu}. A particularly striking feature of these solutions is that they contain no central physical singularity, a fact that strongly suggests that shape dynamic black holes do not suffer from an information loss paradox as general relativistic black holes do.  The CPT invariance of electrovac black hole solutions with a high degree of symmetry presents an interesting possible connection to the standard model of particle physics. While shape dynamic black holes have some promising features, there are still some important open questions that need to be addressed, not least of which is whether such solutions can form from collapse, a problem which is already being explored and which promises to shed light on the known eternal solutions. Collapse models are also an important step in understanding two features of shape dynamic black holes which are currently poorly understood: First, what does an observer see as she passes through the horizon of a shape dynamic black hole? And second, do shape dynamic black holes evaporate via Hawking radiation? 

The first of these questions arises as a result of the fact that these solutions do not form well-defined spacetime geometries on the horizon. As a result, the timelike component of the geodesic equation is ill-defined at the horizon, and the trajectories of observers that would ordinarily be expressed as timelike geodesics con no longer be described in this manner. This might be resolved by defining timelike geodesics connecting points in the interior and exterior at different times as piece-wise smooth curves that are timelike geodesics in the interior and exterior regions and that minimize the proper time. In this way the resulting degenerate spacetime becomes geodesically complete, although the geodesics defined in this manner have a discontinuity in their tangent vector at the horizon. While this seems like a reasonable definition, it remains troubling that the trajectories of observers cannot be determined locally as the solution of a second order differential equation that is well-defined everywhere.

The second question was partially addressed in \cite{Vasu} where it was noted that since shape dynamic black holes agree with general relativistic black holes in the exterior region it may be possible to import whole-cloth certain derivations of Hawking radiation that make no reference to the interior region. This can be done by solving the Klein-Gordon equation in the background of the exterior black hole geometry, quantizing the Klein-Gordon field, and relating the ingoing modes at past null infinity to ingoing modes at the horizon and outgoing modes at future null infinity via Bogoliubov transformation \cite{Lambert}. Still, it would be preferable to derive Hawking radiation in the context of a collapse model in order to obtain a more concrete physical picture of the dynamics of black hole formation and evaporation.  

Rindler space provides a simple example of a solution of shape dynamics that might be used to study black hole thermodynamics via the Unruh effect. Moreover, all of the black hole solutions of shape dynamics share the same global causal structure as Rindler space, so this simple example might be used as a toy model to study more complicated questions about shape dynamic black holes. 

The singular parity horizon encountered in the case of the Bonner solution clarifies the role of the assumption of global hyperbolicity in shape dynamics, and suggests a general chronology protection mechanism in shape dynamics. One can assume global hyperbolicity from the outset and then solutions with closed timelike curves are automatically excluded, but there remain solutions of general relativity which admit maximal slicing outside of the chronological Cauchy horizon. The procedure outlined in the case of the Bonner spacetime explains how these two facts can be reconciled. By making a singular coordinate transformation that makes the spatial metric invertible across the Cauchy horizon, one obtains a global solution of shape dynamics in which closed timelike curves do not form, and which contains closed null curves only on a singular sub-surface of measure zero. If one demands continuity of the phase space variables then these solutions must be discarded, since the metric and momentum are divergent there and the singularity in the square of the Cotton tensor ensures that there is no spatial diffeomorphism which can remove these divergences.

Clearly, much work remains in understanding the physical differences between shape dynamics and general relativity, but the solutions presented in this work suggest that parity horizons, or some dynamical generalization thereof my be of great utility in characterizing the possible global differences between the two theories that can arise.
\begin{appendices}
\chapter{Conformally Reduced Connection and Curvature of Kerr Metric in Prolate Spheroidal Coordinates}
The non-zero connection coefficients associated with the metric \eqref{new gauge} are given by
\begin{eqnarray}\label{connection}
\Gamma^{\phi}_{\mu\phi} &=& \frac{1}{2}(\ln\Psi)_{,\mu} \nonumber \hspace{32.27 pt}
\Gamma^{\phi}_{\theta\phi} = \frac{1}{2}(\ln\Psi)_{,\theta} \nonumber \\
\Gamma^{\mu}_{\phi\phi} &=& -\frac{1}{2}\Psi_{,\mu} \hspace{44.27 pt} 
\Gamma^{\theta}_{\phi\phi} = -\frac{1}{2}\Psi_{,\theta} 
\end{eqnarray}

from which we obtain the components of the  Ricci tensor
\begin{eqnarray}\label{Ricci}
R_{\mu\mu} &=& -\frac{1}{2}(\ln\Psi)_{,\mu\mu} - \frac{1}{4}[(\ln\Psi)_{,\mu}]^2 \nonumber \\
R_{\theta\theta} &=& -\frac{1}{2}(\ln\Psi)_{,\theta\theta} - \frac{1}{4}[(\ln\Psi)_{,\theta}]^2 \nonumber \\
R_{\mu\theta} &=& -\frac{1}{2}(\ln\Psi)_{,\mu\theta} - \frac{1}{4}(\ln\Psi)_{,\mu}(\ln\Psi)_{,\theta} \\
R_{\phi\phi} &=& -\frac{1}{2}(\Psi_{,\mu\mu} +\Psi_{,\theta\theta}) + \frac{1}{4\Psi}[(\Psi_{,\mu})^2 + (\Psi_{,\theta})^2] \nonumber \\
R_{\mu\phi} &=& R_{\theta\phi} = 0 \nonumber \\ \nonumber
\end{eqnarray}

and the Ricci scalar 
\begin{equation}\label{R}
R = \frac{1}{2\Psi^2}[(\Psi_{,\mu})^2 + (\Psi_{,\theta})^2] - \frac{1}{\Psi}(\Psi_{,\mu\mu} + \Psi_{,\theta\theta}) \bigskip \\
\end{equation}

Using (\ref{connection}), (\ref{Ricci}), and (\ref{R}) we can construct the Cotton-York tensor
\begin{equation}\label{Cotton-York}
C^{ij} = \epsilon^{ikl}\left(R^j_{\hphantom{j}l;k} - \frac{1}{4}\delta^j_{\,\,l} R_{,k} \right)
\end{equation}

which is by construction symmetric, traceless and transverse. Putting (\ref{connection}), (\ref{Ricci}), and (\ref{R}) into (\ref{Cotton-York}) yield the components of the Cotton-York tensor 
\begin{eqnarray}\label{CY components}
C^{\mu\mu} &=& C^{\mu\theta} = C^{\theta\theta} = C^{\phi\phi} = 0 \nonumber \\
C^{\mu\phi}  &=& \frac{1}{\Psi}R_{\phi\phi,\theta} -\frac{1}{4}R_{,\theta} \\
C^ {\theta\phi} &=& -\frac{1}{\Psi}R_{\phi\phi,\mu} + \frac{1}{4}R_{,\mu} \nonumber
\end{eqnarray}

from which one can form the scalar density 
\begin{eqnarray}\label{CY squared}
C^2 := C^{ij}C_{ij} &=& 2\Psi\left[ (C^{\mu\phi})^2 + (C^{\theta\phi})^2 \right] \nonumber \\ &=& \frac{1}{4\Psi^2}\left[ \Psi_{,\mu}\Psi_{,\mu\mu} + \Psi_{,\theta}\Psi_{,\theta\theta} + (\Psi_{,\mu} + \Psi_{,\theta})\Psi_{,\mu\theta} \right] \nonumber \\ 
&\hphantom{=}& -\frac{1}{4\Psi}\left( \Psi_{,\mu\mu\mu} + \Psi_{,\mu\mu\theta} + \Psi_{,\mu\theta\theta} + \Psi_{,\theta\theta\theta} \right).
\end{eqnarray}
\pagebreak

\chapter{Poisson Bracket Between Horizon Hamiltonian and Conformal Constraint}
 In order to calculate the bracket $\{H_{hor}(N), \pi(\rho)\}$, it is first necessary to
transform $H_{hor}(N)$ back into a volume integral:
\begin{gather}
H_{hor}(N)
=
\int_{\partial\Sigma}d^2y\sqrt{\sigma}\frac{\kappa}{8\pi}m^{2\hat{\phi}}
\nonumber\\ 
=
\frac{1}{8\pi}\int_{\partial\Sigma}d^2y\sqrt{\sigma}n^i\left(\nabla_iNe^{2\hat{\phi}}\right) \label{bulk
H hor}\\
=
\frac{1}{8\pi}\int_{\Sigma}d^3x\sqrt{q}\hspace{.1cm}\nabla^i\left(\nabla_iNe^{2\hat{\phi}}\right) \nonumber
\end{gather}

\noindent Since the covariant derivative $\nabla_i$ depends on the metric, it is
not immediately clear how to treat it under the Poisson
bracket. Usually, this can be taken care of using repeated integration
by parts and making use of compactness or boundary conditions. Here
things are complicated by the presence of the extra boundary at the
horizon, so this is not the most convenient way to compute the
bracket. Rather, I make use of the lapse-fixing equation obtained from
phase-space reduction:
\begin{equation}\label{LFE}
\sqrt{q}\,\nabla^i\left(\nabla_i Ne^{2\hat{\phi}}\right) = e^{-6\hat{\phi}}\frac{N}{\sqrt{q}}G_{ijkl}\pi^{ij}\pi^{kl}.
\end{equation}

\noindent Inserting \eqref{LFE} into \eqref{bulk H hor} and acting with the conformal constraint $\pi(\rho) := \int_{\Sigma}d^3x\rho q_{ij}\pi^{ij}$, I obtain:
\begin{gather}
\{H_{hor}(N), \pi(\rho)\} = \bigg\{\left(
\frac{1}{8\pi}\int_{\Sigma}d^3x\hspace{.1cm}e^{-6\hat{\phi}}\frac{N}{\sqrt{q}}G_{ijkl}\pi^{ij}\pi^{kl}
\right),\left( \int_{\Sigma}d^3x^{\prime}\rho
  q_{mn}\pi^{mn}\right)\bigg\} \nonumber \\
=
\frac{1}{8\pi}\int_{\Sigma}d^3x\int_{\Sigma}d^3x^{\prime}
\bigg[
\left(e^{-6\hat{\phi}}N\pi^{ij}\pi^{kl}\right)(x)\left(\rho
q_{mn}\right)(x^{\prime})
\bigg\{\left(\frac{G_{ijkl}}{\sqrt{q}}\right)(x)
,\pi^{mn}(x^{\prime})\bigg\} \label{bracket1}\\
+ \left(  e^{-6\hat{\phi}}\frac{N}{\sqrt{q}}G_{ijkl}\right)(x)\left(\rho\pi^{mn}\right)(x^{\prime})
\bigg\{\left(\pi^{ij}\pi^{kl}\right)(x) ,q_{mn}(x^{\prime})\bigg\}
\bigg]. \nonumber
\end{gather}

\noindent By direct computation, I find
\begin{gather}
\pi^{ij}\pi^{kl}q_{mn}\bigg\{\left(\frac{G_{ijkl}}{\sqrt{q}}\right)(x)
,\pi^{mn}(x^{\prime})\bigg\} =
-\frac{G_{ijkl}}{2\sqrt{q}}\pi^{ij}\pi^{kl}\delta(x-x^{\prime}) \\
\frac{G_{ijkl}}{\sqrt{q}}\pi^{mn}\bigg\{\left(\pi^{ij}\pi^{kl}\right)(x)
,q_{mn}(x^{\prime})\bigg\} = -\frac{2G_{ijkl}}{\sqrt{q}}\pi^{ij}\pi^{kl}\delta(x-x^{\prime})
\end{gather}

\noindent which finally yield the result
\begin{gather}\label{last bracket}
\{H_{hor}(N), \pi(\rho)\} = -\frac{3}{16\pi}\int_{\Sigma}d^3x\,
\frac{N\rho}{\sqrt{q}}e^{-6\hat{\phi}} G_{ijkl}\pi^{ij}\pi^{kl} =
-\frac{3}{2}H_{hor}(\rho N).
\end{gather}

\pagebreak

\chapter{Identification of the Constant in the Rindler Lapse Function as Surface Gravity}
The general solution to Laplace's equation in Cartesian coordinates can be written:
\begin{gather}
\nonumber
N(x,y,z) = \sum\limits_{n_y,n_z =0}^{\infty}\left(A_{x,n_y,n_z} e^{\frac{x}{2\pi}\sqrt{n_y^2 -n_z^2}} + B_{x,n_y,n_z} e^{-\frac{x}{2\pi}\sqrt{n_y^2 -n_z^2}}\right) \nonumber\\ \times \left(A_{y,n_y}\sin\left(\frac{n_y y}{2\pi}\right)+B_{y,n_y}\cos\left(\frac{n_y y}{2\pi}\right)\right) \label{rectangular harmonics}\\ \times \left( A_{z,n_z}\sin\left(\frac{n_z z}{2\pi}\right)+B_{z,n_z}\cos\left(\frac{n_z z}{2\pi}\right)\right). \nonumber
\end{gather}
\noindent Imposing $N(0,y,z) = 0$, one obtains $A_x + B_x = 0$, so the solution \eqref{rectangular harmonics} becomes:
\begin{gather}
N(x,y,z) = \sum\limits_{n_y,n_z =0}^{\infty}2A_{x,n_y,n_z}\sinh\left(\frac{x}{2\pi}\sqrt{n_y^2 -n_z^2}\right) \nonumber \\
\times\left(A_{y,n_y}\sin\left(\frac{n_y y}{2\pi}\right)+B_{y,n_y}\cos\left(\frac{n_y y}{2\pi}\right)\right) \label{rectangular harmonics + BC1}\\
\times\left(A_{z,n_z}\sin\left(\frac{n_z z}{2\pi}\right)+B_{z,n_z}\cos\left(\frac{n_z z}{2\pi}\right)\right). \nonumber 
\nonumber
\end{gather}
\noindent If one then imposes 
\begin{equation}\label{firstDer}
\pder{N}{x}\bigg|_{x=0} = \kappa 
\end{equation}

\noindent the only allowed values of $n_y$ and $n_z$ are zero, so the A's and B's corresponding to $n_y, n_z \neq 0$ must all vanish, and one obtains
\begin{equation}\label{BC2}
\kappa = \frac{c_{n_y,n_z}}{2\pi}\sqrt{n_y^2 -n_z^2}\bigg|_{n_y=n_z=0}
\end{equation}

\noindent where $c_{n_y,n_z} = 2A_{x,n_y,n_z}B_{y,n_y}B_{z,n_z}$. Solving \eqref{BC2} for $c_{n_y,n_z}$ and inserting the result back into \eqref{rectangular harmonics + BC1} gives
\begin{equation}\label{limit1}
N(x) = \frac{2\pi\kappa}{\sqrt{n_y^2 -n_z^2}}\sinh\left(\frac{x}{2\pi}\sqrt{n_y^2 -n_z^2}\right)\bigg|_{n_y=n_z=0}
\end{equation}

\noindent Clearly, this equation must be interpreted as a limit, which can be evaluated using L'Hospital's rule for indeterminate forms:
\begin{eqnarray}\label{limit 2}
N(x) = \lim_{a \to 0}\frac{2\pi\kappa}{a}\sinh\left(\frac{ax}{2\pi}\right)
= \lim_{a \to 0}\frac{2\pi\kappa}{1}\frac{x}{2\pi}\cosh\left(\frac{ax}{2\pi}\right) = \kappa x.  
\end{eqnarray}
\end{appendices}

\bibliography{DisBib2}

\end{document}